\documentclass[rmp,aps,twocolumn]{revtex4}
\usepackage{graphics}
\usepackage{epsfig}

\begin{document}
\title{Looking at a soliton through the prism of optical supercontinuum}

\author{Dmitry V. Skryabin}\email{d.v.skryabin@bath.ac.uk}
\author{Andrey V. Gorbach}
\affiliation{Centre for Photonics and Photonic Materials, Department
of Physics, University of Bath, Bath, BA2 7AY, United Kingdom}

\begin{abstract}
A traditional view on  solitons in optical fibers as  robust particle-like
structures suited for information transmission has been significantly altered and broadened
over the past decade, when solitons have been found to play the major role
in generation of octave broad supercontinuum spectra in photonic-crystal and other types of optical
fibers. This remarkable spectral broadening is achieved through complex processes
of dispersive radiation being scattered from, emitted and transformed by  solitons. Thus solitons
have emerged as the major players in nonlinear frequency conversion in optical fibers.
Unexpected analogies
of these  processes have been found with  dynamics  of ultracold atoms and ocean waves.
This colloquium focuses on recent understanding and  new
insights into physics of  soliton-radiation interaction
and supercontinuum generation.

\end{abstract}

%\date{May 2001}
\maketitle
\tableofcontents

\section{Introduction}

Interplay of dispersion and nonlinear self-action in wave dynamics
has been at the focus of attention across many branches of physics
since the middle of the past century after the seminal
Fermi-Pasta-Ulam work on heat dissipation in solids.
This work has been quickly followed by the discovery of many
nonlinear wave equations integrable with the inverse scattering technique (IST.
Solitons is a particular class of  localized and remarkably robust solutions found with the IST technique.
Soliton studies have quickly become a subject of its own and have soon developed
far beyond the initial subset of integrable models.
Localised nonlinear waves in the non-integrable models are  often called
solitary waves to distinguish them from IST solitons.
However,  the use of the original term 'soliton' has now
spread widely into non-integrable cases.
Solitons have been experimentally observed and studied theoretically
in fluid dynamics, mechanical systems,
condensed matter and notably in nonlinear optics.
Historic and  scientific accounts of these developments can be found, e.g., in
the excellent book by Alwyn Scott \cite{Scott1999}.

Optical solitons in fibers \cite{Hasegawa1973,MOLLENAUER1980}, have been most extensively
researched as potential information carriers \cite{Agrawal2001,Mollenauer2006}.
With this view in mind  solitons can be treated as particle-like objects
and their dynamics can be conveniently reduced to the Newton-like equations for
 the soliton degrees of freedom, such as, e.g.
position and phase \cite{Agrawal2001,Mollenauer2006,GORSHKOV1981,KAUP1978}.
Ability of perturbed solitons to emit dispersive radiation
is a property vividly expressing their wave nature. Radiation emission by solitons  has been
well known in 20th century  \cite{WAI1990,Karpman1993,Akhmediev1995,Kivshar1989},
however, it was considered at that time  as
something that to the large degree undermines  usefulness of the soliton
concept. This vision has changed dramatically around and after 2000,
when the first experimental observation of an octave wide spectral
broadening in photonic crystal fibers was reported by Ranka et al. \cite{Ranka2000}
and subsequently reproduced in dozens of labs. This effect has
become known as the fiber supercontinuum, see, e.g.,
\cite{Russell2006,Dudley2006,Knight2007,Dudley2009} for recent reviews, and Fig. \ref{skr0}.
The above reviews  discuss  impressive applications of  supercontinuum for
frequency comb generation, in metrology, spectroscopy and imaging.

\begin{figure}
\centering
\caption{(color online) Photograph of supercontinuum generated by a 800nm femtosecond pulse
in a photonic crystal fiber and projected onto a screen.
(http://www.bath.ac.uk/physics/groups/cppm/).}
%The total spectrum extends from $\sim 400$ to $\sim 2000$nm.}
\label{skr0}
\end{figure}

Already after
first experiments on generation of  supercontinuum in fibers
it has become obvious that
solitons are the major players in this process. Spectrally they dominate
in the  infrared, where the group velocity dispersion (GVD) is typically anomalous.
The visible part of  supercontinua most often spans through the range of normal GVD and
is  associated with dispersive radiation. Crucially,  the dispersive
radiation and solitons overlap in the time domain and hence
interact by means of the Kerr and Raman nonlinearities.

What was  overlooked in the research on fiber solitons in the pre-supercontinuum
era is the fact that interaction of solitons with dispersive waves and the associated frequency
conversion processes can be efficient and practically important. Not only experiments,
but also theory of the soliton-radiation interaction
was  not  developed  much beyond the spectrally narrow results following from
the integrable Nonlinear Schrodinger (NLS) equation, where the solitons
are largely insensitive to the interaction with other waves \cite{ZAKHAROV1972,KUZNETSOV1995,Kivshar1989}.
Our primary aim here is to explain the frequency conversion effects resulting
from the soliton-radiation interaction and leading to supercontinuum generation in optical fibers,
and to discuss few other fascinating soliton related effects, which have stem from the supercontinuum
research.

\section{Modeling of supercontinuum}
\begin{figure}
\centering
\includegraphics[width=7cm]{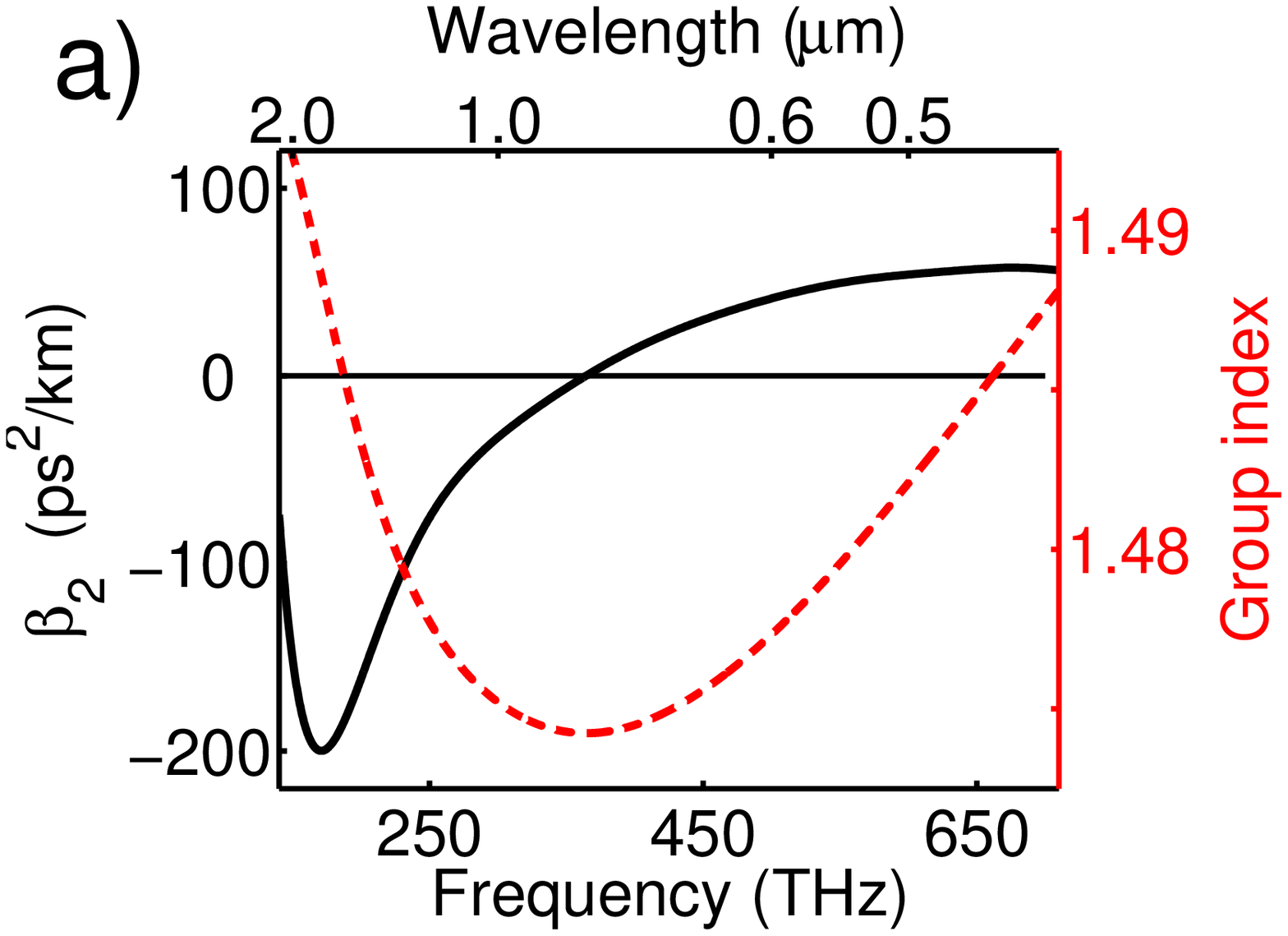}
\includegraphics[width=7cm]{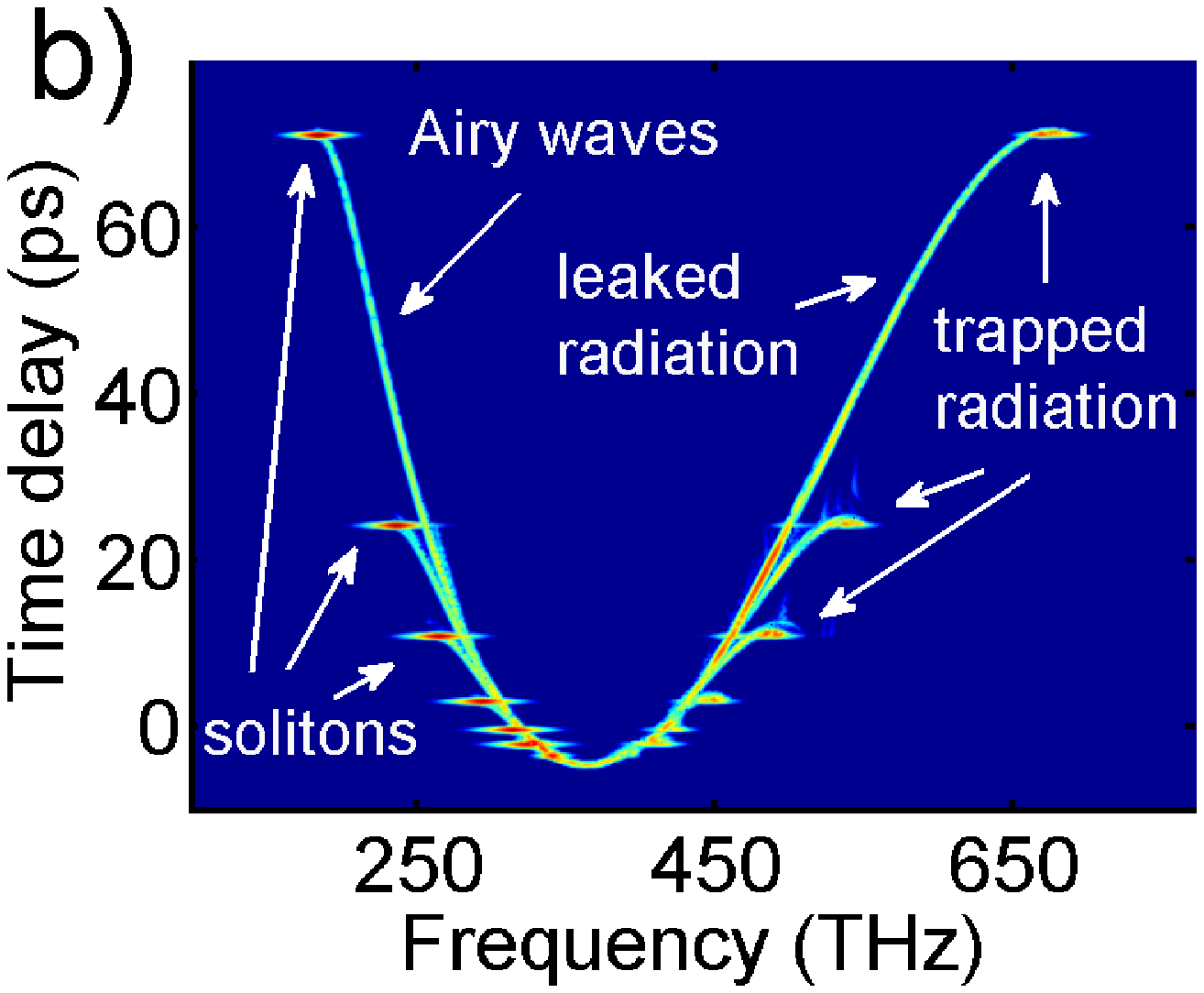}
\caption{(color online) (a) Group index, that is, speed of light divided by the
group velocity, $n_g=c\partial_{\delta}k$, (dashed  line) and GVD
$\beta_2=\partial_{\delta}^2k$ (full line) typical for photonic-crystal
fibres used in supercontinuum experiments. The zero GVD point is at
$\simeq 790$nm ($380$THz) and the normal GVD range is for wavelengths $<790$nm.
(b) XFROG spectrogram showing simultaneous frequency
and time domain pictures of the supercontinuum from Fig. \ref{skr2}
at $z=1.5$m. Pump wavelength is $850$nm ($350$ THz). } \label{skr1}
\end{figure}

Talking about supercontinuum generation and soliton-radiation interaction
we consider a fiber with sufficiently high  nonlinearity  and the
zero  GVD point close to the pump wavelength. These conditions are found in fibers with silica
cores of few ($\sim$ $1-5$) microns in diameter, which can be, e.g.,
either photonic crystal  or tapered fibers pumped with a variety of
sources. The later include femtosecond pulses of   mode-locked Ti:sapphire
lasers with wavelength around $800$nm \cite{Ranka2000} and nano-second pulses from
microchip lasers close to $1\mu$m \cite{Stone2008}. An important aspect of the
dispersion profile allowing to achieve wide supercontinua is the
normal GVD extending towards  shorter ('bluer') wavelengths, see
Fig. \ref{skr1}(a). Spectral broadening with the opposite
GVD slope has also been studied \cite{Efimov2004,Harbold2002}.

The widely accepted model well reproducing
experimental measurements of supercontinuum is
\begin{eqnarray}
\label{eq1} &&\partial_zA=i k(i\partial_t)A+\\
\nonumber &&i\gamma(1-\theta)|A|^2A+i\gamma\theta
A\int_0^{\infty}dt'R(t')|A(t-t')|^2.
\end{eqnarray}
Here $z$ is the propagation length and $t$ is the time measured
in the reference frame moving with the group velocity at the pump
frequency.
$\gamma\sim 0.01$/W/m is the nonlinear parameter.
$\theta=0.18$ measures the strength of the Raman nonlinearity with respect to
the Kerr one and $R(t)$ is the Raman response function \cite{Gorbach2006,Agrawal2001}.
$A$ is the modal amplitude, which carries all the spectrum we are
interested in. $k(i\partial_t)$ is the dispersion operator, usually represented
as a polynomial of the order $N$ in $i\partial_t$.
Dispersion profile of a linear fiber is recovered  by taking
$A=e^{ik(\delta)z-i\delta t}$, where
$k(\delta)=\sum_{n=2}^N{1\over n!}\beta_n\delta^n$,
which is a polynomial fit (not a spectrally local Taylor expansion) of the fiber dispersion.
Positive $\delta$'s correspond to the
 absolute frequencies larger than the pump frequency.
 GVD coefficient $\beta_2$ is calculated as $\beta_2=\partial_{\delta}^2k$.
$\beta_2=0$  determines location of the zero GVD frequencies.

Eq. (\ref{eq1}) includes  three  ingredients:
\begin{itemize}
\item {\em dispersion} (1st term in the righthand side)
\item {\em nonlinear phase modulation}  due to instantaneous Kerr nonlinearity (2nd
term)
\item {\em Raman scattering} (3rd term)
\end{itemize}
Taken separately  these effects are well known. Notably,
the last two lead to generation of new spectral content  acting on
their own. However,   conversion into discrete Raman side-bands
 does not manifest itself in the typical
supercontinuum experiments. Also self- (SPM) and cross- (XPM) phase modulations
(as particular cases of the generic nonlinear phase modulation) generate spectra, which are
relatively narrow \cite{Agrawal2001}, and can
not explain fully developed supercontinua. It is only when dispersion comes
into play  and all three
of the above effects act together in symphony, then the qualitatively new phenomenon
of supercontinuum generation occurs. A typical supercontinuum
experiment with the femtosecond $800$nm pump leads to spectra
covering range from $400$nm to $2000$nm after propagation in
a $\sim 1$m long photonic crystal fiber \cite{Ranka2000,Gu2003,Wadsworth2002}.
Figs. \ref{skr1}(b) and \ref{skr2} show numerical
modelling of this process using Eq. (\ref{eq1}).
\begin{figure}
\centering
\caption{(color online) Numerical simulation of supercontinuum generation
in a photonic-crystal fibre pumped with 200-fs pulses at 850 nm and having 6-kW peak
power.  The fiber dispersion as in Fig. \ref{skr1}(a).
One can see 3 stages in the supercontinuum expansion.
Symmetric spectral broadening due to SPM happens over first $~5$cm. Antisymmetric spectral broadening
due to soliton fission accompanied by the emission of the resonant
Cherenkov radiation and soliton-radiation
interaction both develop between $\sim 5$cm and $\sim 15$cm. After that the dynamics of
the short wavelength edge is determined by the  radiation trapping.
} \label{skr2}
\end{figure}

Results obtained using Eq. (\ref{eq1}) have been directly compared with many
experimental measurements of femtosecond pulse propagation
in photonic crystal fibers in both time and frequency domains and
excellent  agreement  has been established
\cite{Gu2003,Skryabin2003,Efimov2004,Efimov2005a,Efimov2006,Gorbach2006}.
Probably the most convenient way of representing the data for such comparisons is
using  the cross-correlation frequency resolved optical
gating (XFROG) spectrograms \cite{Gu2003,Hori2004,Efimov2004}.
The XFROG spectrogram is the fourier transform of a product of the signal field $E$
with the reference pulse $E_{r}$ delayed by time $\tau$:
$I_{XFROG}(\tau,\delta)=|\int_{-\infty}^{\infty}E_{r}(t-\tau)E(t)e^{-it\delta}dt|^2$.
$E_r$ is usually a pump pulse  and the product of the two fields
is generated by the sum frequency process in a $\chi^{(2)}$ crystal.

As with any model, Eq. (\ref{eq1}) has limits to its applicability.
In particular, it is not applicable to describe
sharp field variations happening over  few femtoseconds
and less. However, so far the role of such ultrashort features in
typical fiber supercontinuum experiments has not been revealed and
the octave spectral broadening happens through nonlinear
interactions of dispersive waves and solitons
which both are well described by Eq. (\ref{eq1}).
In particular,  formation of the coupled  soliton-radiation states
with the continuously blue shifting radiation component and the red shifting
soliton  plays  a key  role in supercontinuum expansion
\cite{Gorbach2007a,Gorbach2007c,Cumberland2008,Travers2009,Stone2008,Kudlinski2008,Hill:09}.

The relative smoothness of a temporal field profile associated with a typical supercontinuum
is one of the reasons why  self-steepening (wave breaking) \cite{Agrawal2001} does
not make a notable quantitative and qualitative impact if included or excluded
from Eq. (\ref{eq1}). Furthermore, going beyond the amplitude model is likely to be needed
for  correct description of  self-steepening, see, e.g., \cite{amiranashvili:063821}.
Therefore we have opted for not taking it into account.
In our experience, the most noticeable disagreements of Eq. (\ref{eq1})
with the experimental measurements are due to omission of the frequency dependent losses.
Other  neglected effects  are possible excitation of
higher order or orthogonally polarized modes, dispersion of the Kerr nonlinearity,
third harmonic generation and noise.
Spontaneous Raman noise and input noise are unavoidably present in
experiments. The spectra averaged over the statistical ensemble may, in some cases,
be better suited for a comparison with  measurements,
where the fine spectral features of a single pulse excitation can be
obscured due to resolution and long (over many pulses) integration time
of a spectrum analyser \cite{Dudley2002,Corwin2003}.
Small to moderate pulse to pulse fluctuations in the generated spectra
should not prevent us from unraveling and understanding of the deterministic
nonlinear processes underlying the supercontinuum generation and the soliton role in it.

\section{Soliton  self-frequency shift and soliton dispersion}
During the first stage of the supercontinuum
development SPM induces spectral broadening and associated chirping
of a femtosecond pulse \cite{Agrawal2001}, which  are  compensated by the anomalous GVD
($\partial_{\delta}^2k<0$). Competition of these processes initiates formation of
multiple solitons from an initial high power pump pulse (soliton
fission), see Fig. \ref{skr2}. Amongst the host of the soliton
related effects, there are two, which are particularly important for us.
These are the soliton self-frequency shift induced by the Raman
scattering \cite{Mitschke1986} and emission of dispersive radiation.
The latter comes from the overlap of the soliton spectrum with
the range of frequencies where the fiber GVD is normal
\cite{WAI1990,Karpman1993,Akhmediev1995}.

Spectra of femto- or pico-second solitons are sufficiently narrow, so
that the Raman gain profile of silica peaking at $13$THz
can be approximated by a straight line rising across the
soliton spectrum from the negative values (damping) at the
short-wavelength end of the soliton spectrum to the positive values
(gain) at the long-wavelength end \cite{Agrawal2001,Luan2006a,Gorbach2008}. The net result on the
soliton is that its spectral center of mass is shifted towards
redder frequencies \cite{Mitschke1986}, which is referred as the soliton
self-frequency shift. The group index $n_g$ (ratio of the vacuum speed  of
light $c$ to the group velocity, $n_g=c\partial_{\delta} k$)
increases with the wavelength providing
that the GVD is anomalous. This leads to continuous negative acceleration (deceleration)
of solitons by the Raman scattering, which is very important for supercontinuum,
but often forgotten in soliton studies.

\begin{figure}
\centering
\includegraphics[width=7cm]{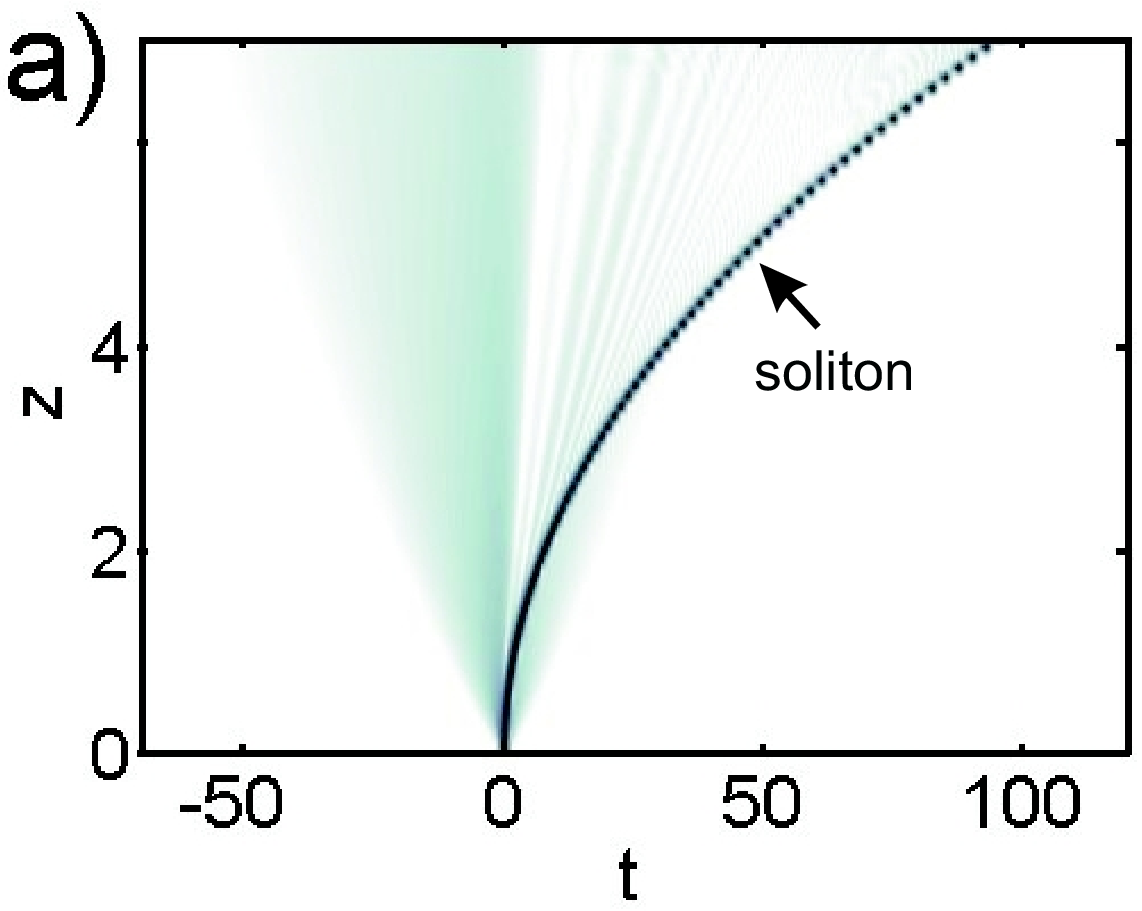}
\includegraphics[width=4cm]{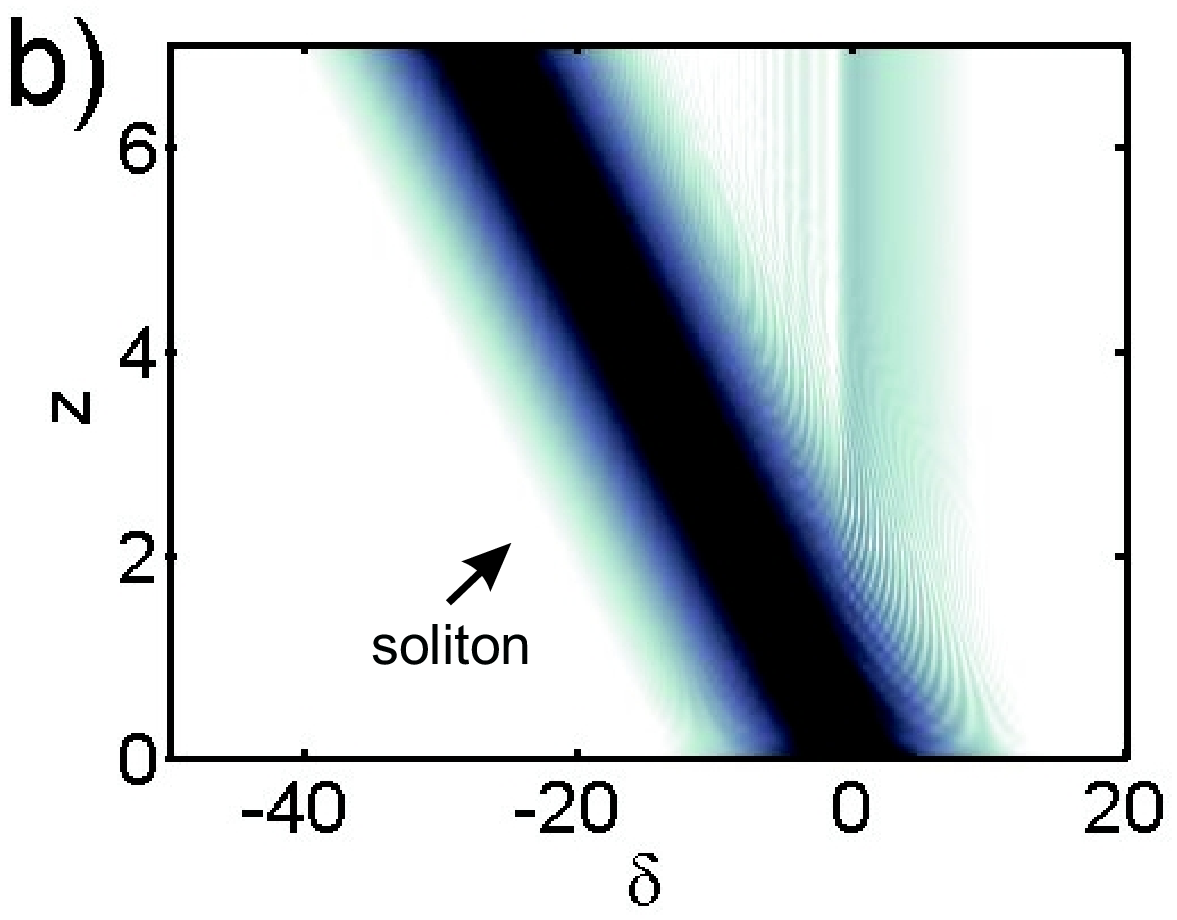}
\includegraphics[width=4cm]{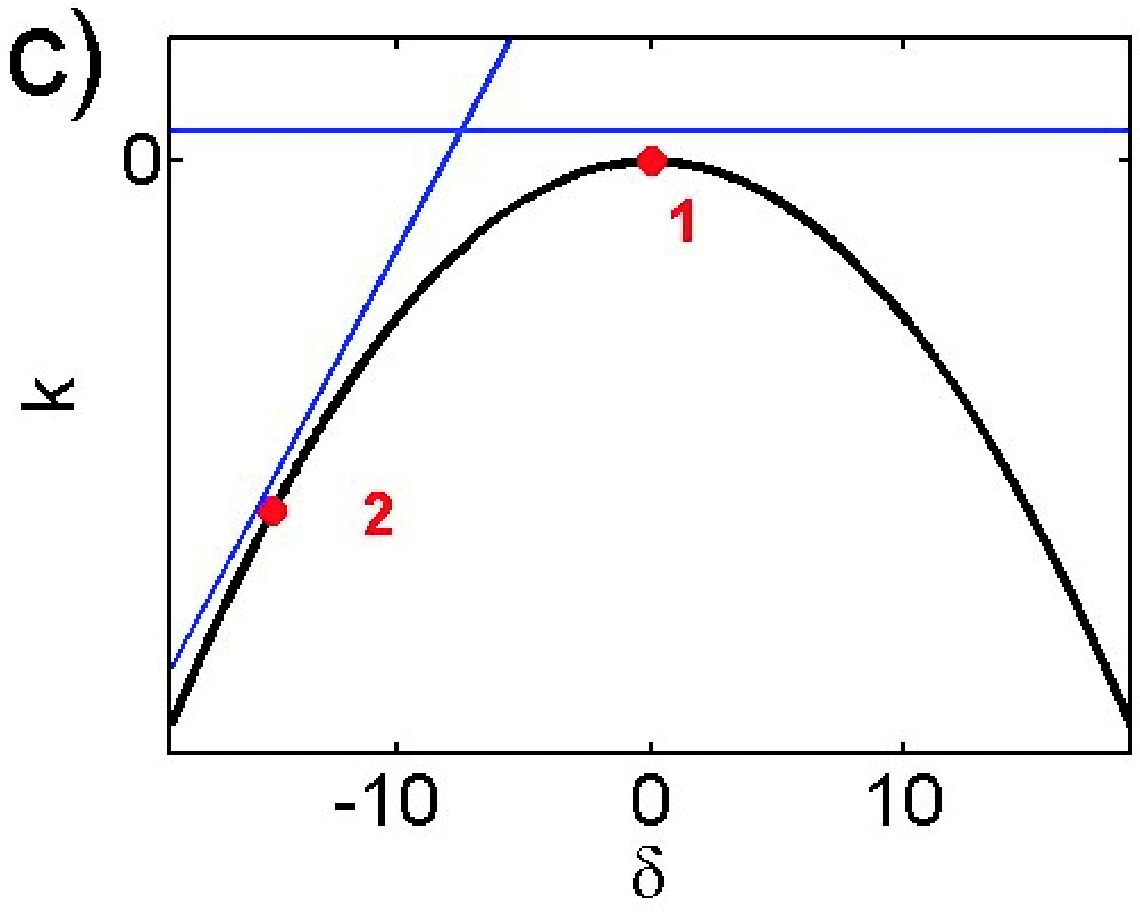}
\caption{Soliton evolution in  $(z,t)$ (a) and
$(z,\delta)$ (b) planes in the presence of the Raman
effect. $k(\delta)={1\over 2}\beta_2\delta^2$. Note that the dispersive waves in (a)
are not reflected by the soliton. (c) shows  phase matching diagrams
between the soliton and dispersive waves (see Eq. (\ref{eq8})). Straight
lines are the soliton wavenumbers $k_s$ for two different frequencies $\delta_s$
(see Eq. \ref{eq6}). Point $1$ indicates the
initial soliton frequency $\delta_s$ and point $2$ is $\delta_s$ at some
distance down the fiber. Parabola shows the dispersion of linear waves
$k(\delta)$. All  units  are dimensionless.} \label{skr3a}
\end{figure}

An approximate soliton solution of Eq. (\ref{eq1})
moving with a constant acceleration
can be derived under the assumption that
the fiber group index varies linearly in frequency
(i.e., $\partial_{\delta }k\sim\delta$ and hence $k\sim\delta^2$).
For the zero initial frequency, this solution  is given by
\cite{Gagnon1990,Gorbach2007c,Gorbach2007b}:
\begin{eqnarray}
\label{eq3}
&& A_s=\psi(t-t_s)\exp\left[-i\delta_s t+i\phi(z)\right],\\
\label{eq4}
&& \psi(t)={\sqrt{2q/\gamma}}~\textrm{sech}\left(\sqrt{2q\over|k''|}~t\right),\\
&& \phi(z)=qz+{1\over 3}k''\delta_s^2z,~T=\int_0^{\infty}tR(t)dt,\\
\label{eq4a} && \delta_s={g_0z\over k''},~k''=\partial_{\delta}^2k<0,~g_0=\frac{32Tq^2}{15}.
\end{eqnarray}
The soliton delay caused by the Raman effect is $t_s=g_0z^2/2$.
$q>0$ is the soliton wavenumber shift proportional
to its intensity.
The soliton trajectory in the $(t,z)$-plane  is a parabola given by $t=t_s$ (see
Fig. \ref{skr3a}(a)).
$\delta_s$ is the soliton frequency, which decreases linearly with $z$
(see Fig. \ref{skr3a}(b)).

Soliton-radiation interaction discussed below is sensitive to the phase matching
conditions. Therefore it is important to derive spectral representation for an
accelerating soliton. The  soliton spectrum is calculated as
\begin{equation}
\tilde A_s(\delta)=\int_{-\infty}^{\infty} dt A_se^{i\delta
t}=e^{i\phi(z)+i t_s[\delta-\delta_s]}\tilde\psi(\delta-\delta_s),
\label{sol}
\end{equation}
where $\tilde\psi(\delta)=\int_{-\infty}^{\infty} \psi(t)e^{i\delta t}dt$ is a real function.
Wavenumbers of the soliton spectral components are
\begin{equation}
\label{eq6}
k_s(\delta)=\partial_z[\phi(z)+[\delta-\delta_s] t_s]
=q+\partial_{\delta}k|_{\delta=\delta_s}[\delta-\delta_s]+k(\delta_s).
\end{equation}
$k_s(\delta)$ is linear in $\delta$, expressing the fact that the soliton is immune to
GVD and hence $\partial_{\delta}^2k_s=0$.
Also it can be seen that $k_s(\delta)$
is actually a tangent to the dispersion of linear waves $k(\delta)={1\over 2}\beta_2\delta^2$ taken
at $\delta=\delta_s$ and  shifted away from it
by the offset $q$. If GVD is anomalous for all the frequencies ($\beta_2<0$)
the soliton spectrum does not  touch the spectrum of linear dispersive waves,
see Fig. \ref{skr3a}(c).

$\psi$, as in Eq. (\ref{eq4}),
is of course an approximate solution, and the Raman effect
forces the soliton to shake off
some radiation, see Fig. \ref{skr1}(b), Fig. \ref{skr3a}(a).
This radiation can be approximated by   Airy
functions \cite{Akhmediev1996,Gorbach2008}.  Airy waves  emitted in a typical supercontinuum setting
are very weak, broad band and have  spectrum almost exclusively belonging to the
anomalous GVD range. The latter is the main reason why they practically do not
interact with solitons and their impact  on the supercontinuum
spectrum is secondary in importance.
\begin{figure}
\centering
\includegraphics[width=7cm]{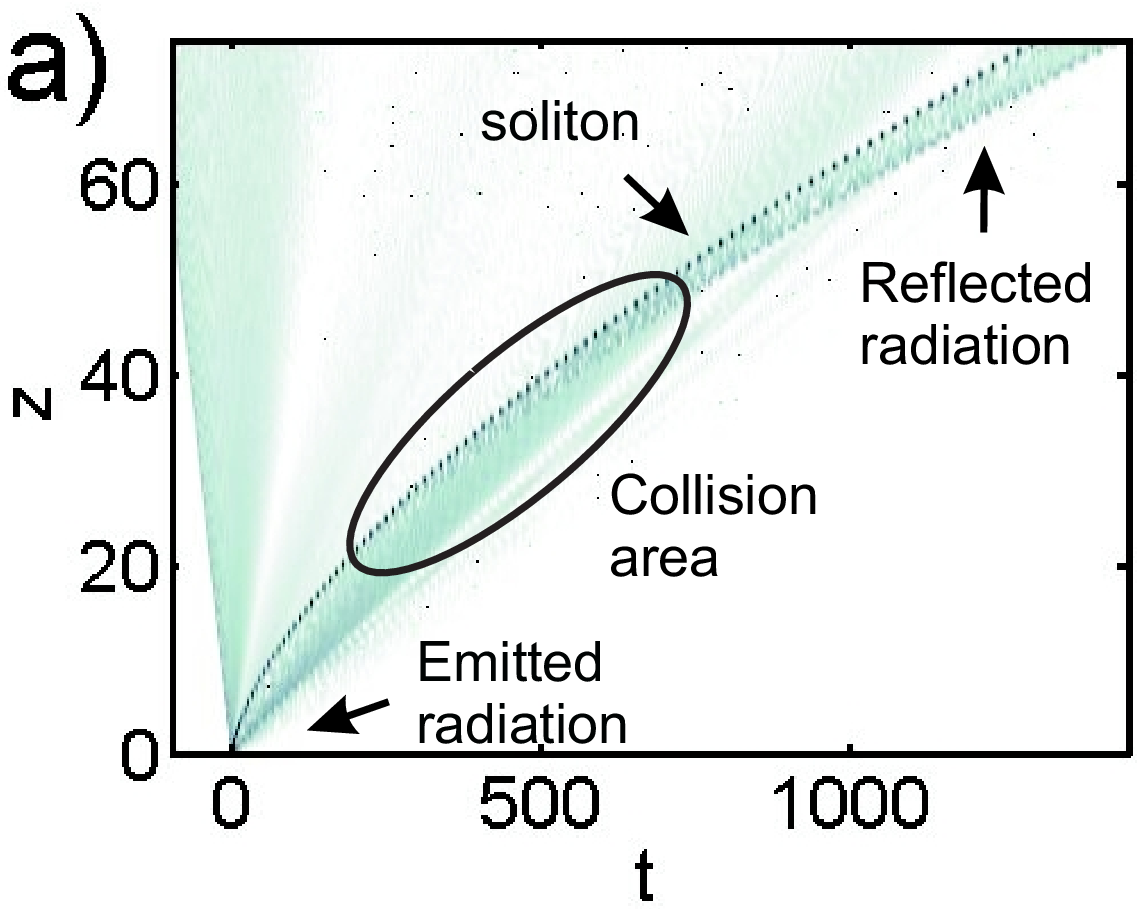}
\includegraphics[width=4cm]{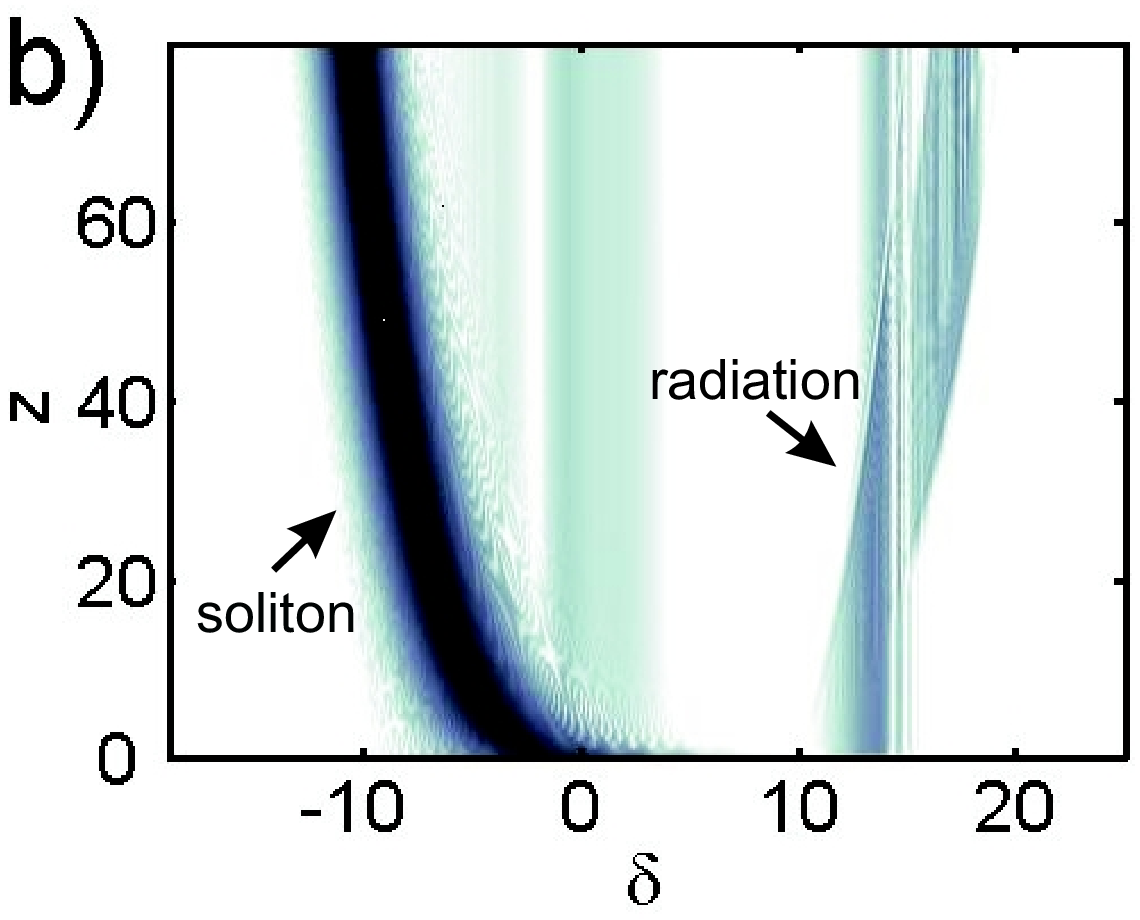}
\includegraphics[width=4cm]{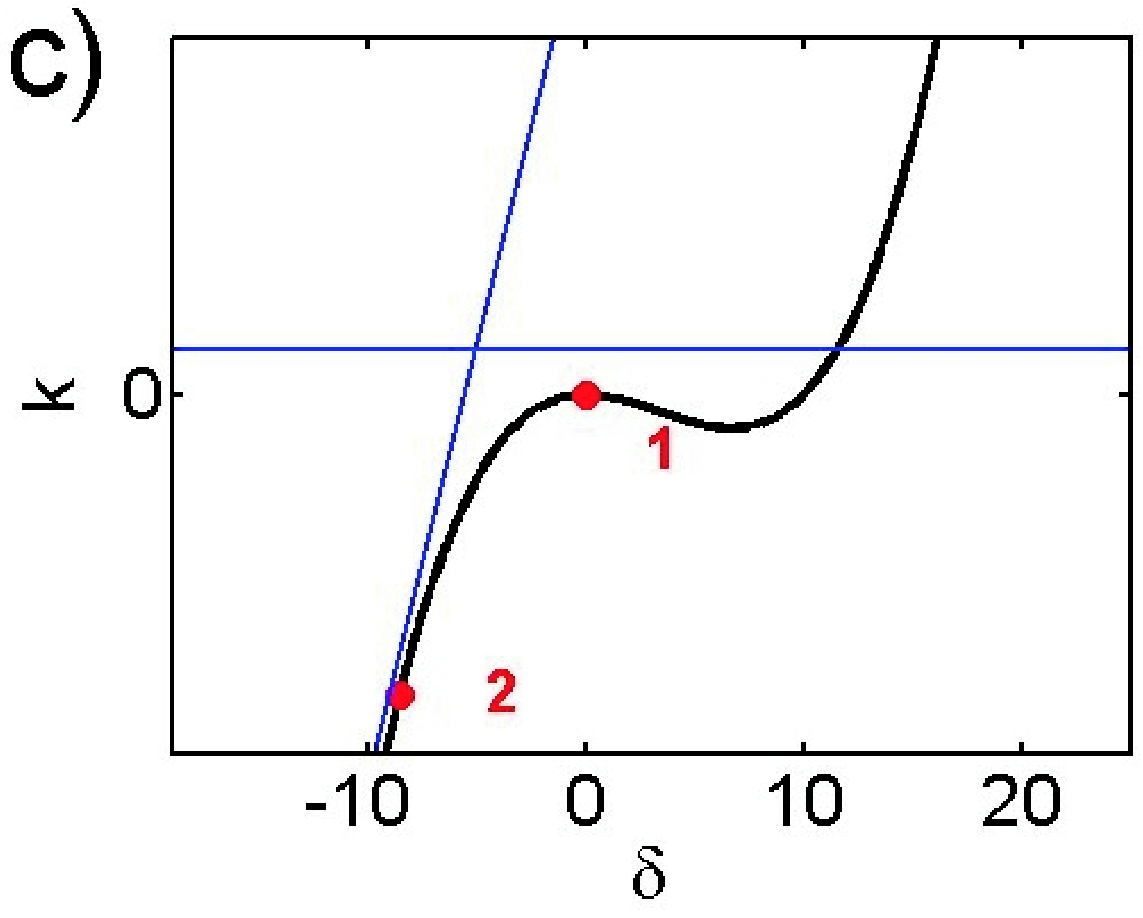}
\caption{The same as Fig. \ref{skr3a}, but with  $\beta_3>0$.
($k(\delta)={1\over 2}\beta_2\delta^2+{1\over 6}\beta_3\delta^3$).
Intersections of the straight lines  with $k(\delta)$, see  (c), correspond
to the Cherenkov resonances. } \label{skr3b}
\end{figure}

\section{Radiation emission by solitons} Solitons dominate the
long-wavelength edge of the supercontinuum (see Figs.1b,2), while the so-called
resonant (or 'Cherenkov') radiation
\cite{WAI1990,Karpman1993,Akhmediev1995,Skryabin2003,Biancalana2004}
comes to mind as one of the  reasons for the spectrum created in the
normal GVD range \cite{Herrmann2002,Gu2003,Cristiani2004}.  Despite
the importance of the resonant radiation identified in the first
efforts to model fiber supercontinuum
\cite{Husakou2001,Herrmann2002}, later it has become clear that it
is only when the initially emitted radiation has a chance to
interact with the solitons over long propagation distance, the
expanding supercontinua observed in experiments can be  reproduced
in modelling
\cite{Skryabin2005,Gorbach2007a,Gorbach2007c,Genty2004b,Genty2005}.
The Raman effect is the key factor  ensuring such interaction.

\begin{figure}
\centering
\includegraphics[width=7cm]{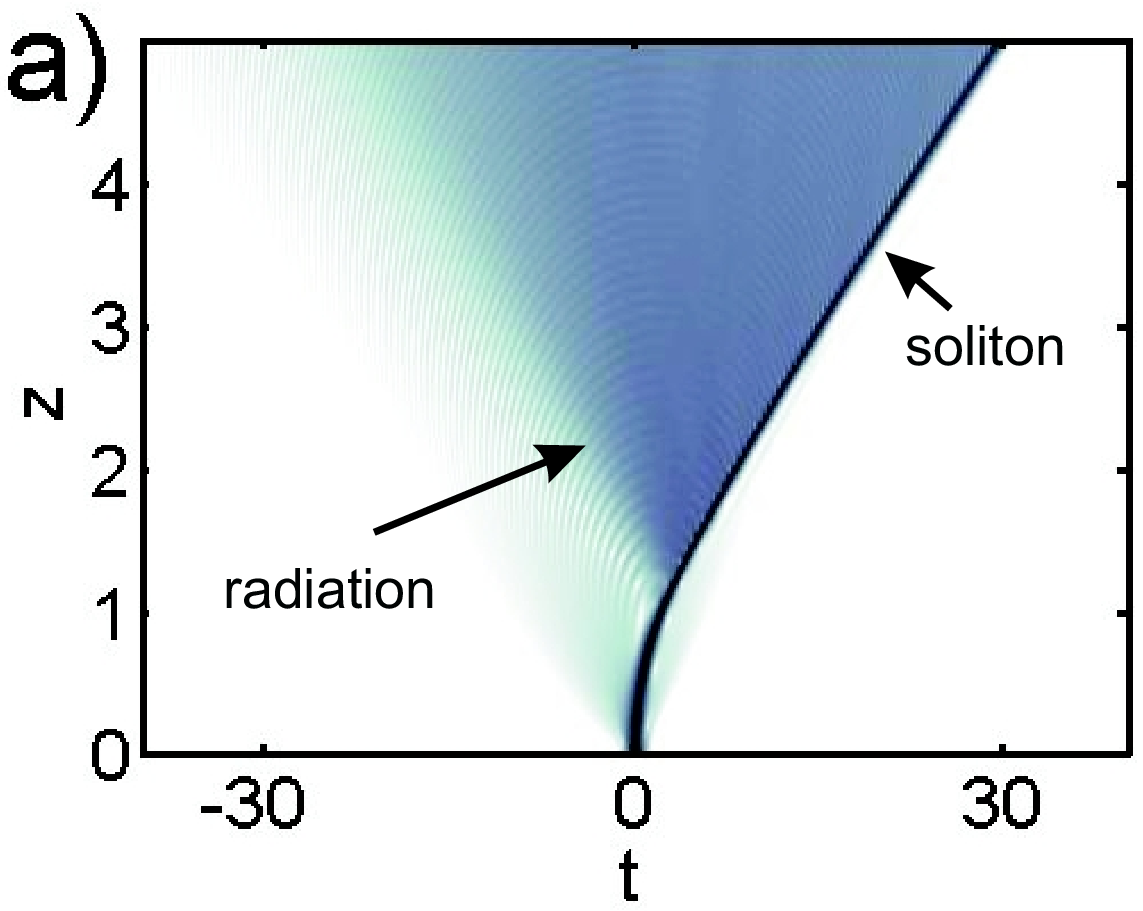}
\includegraphics[width=4cm]{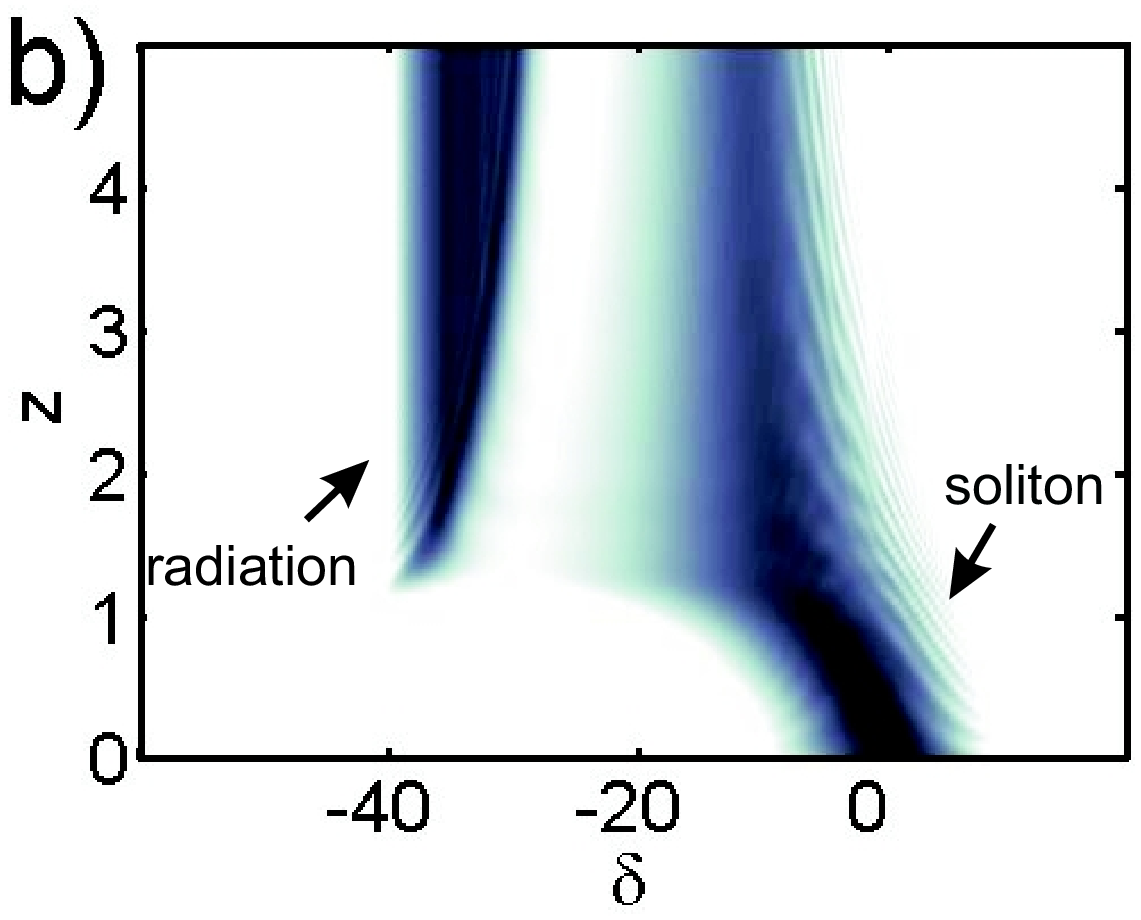}
\includegraphics[width=4cm]{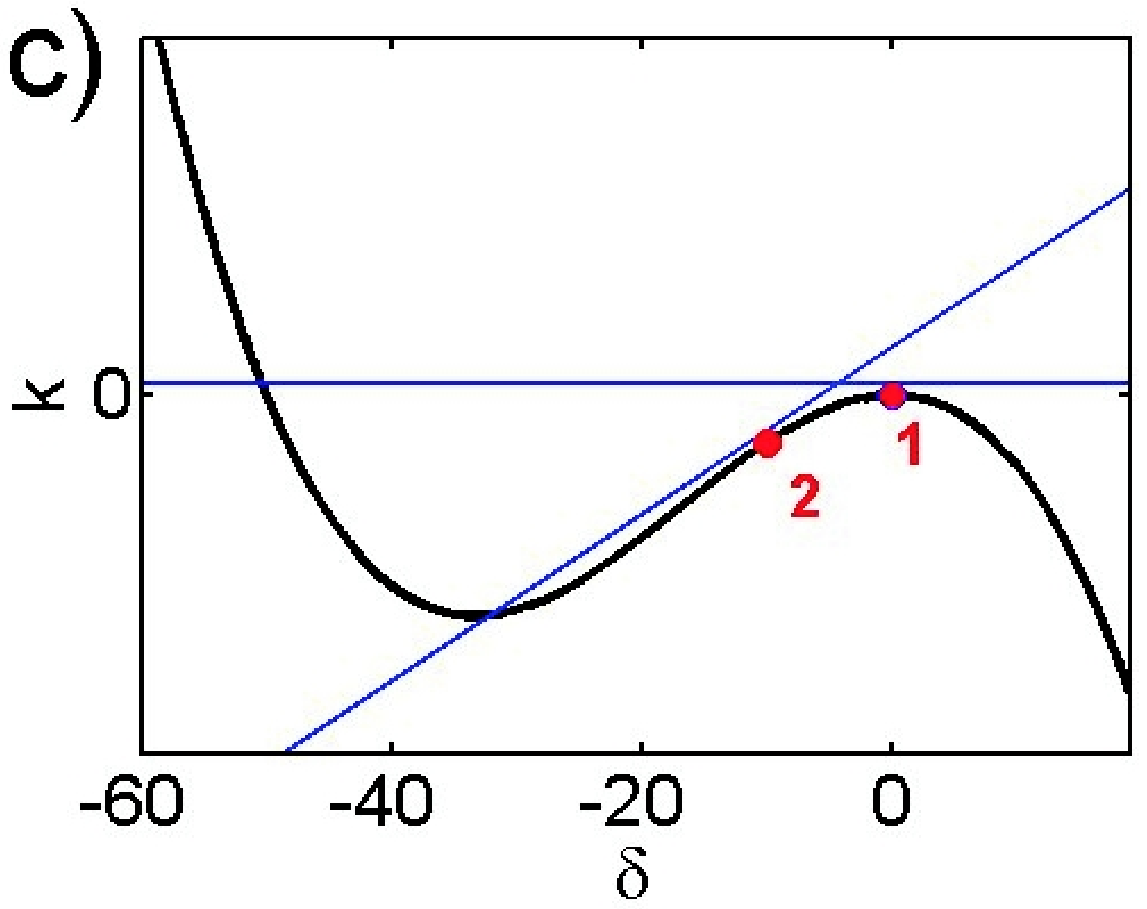}
\caption{The same as Fig. \ref{skr3a}, but with  $\beta_3<0$.
($k(\delta)={1\over 2}\beta_2\delta^2+{1\over 6}\beta_3\delta^3$).
Intersections of the straight lines  with $k(\delta)$, see  (c), correspond
to the Cherenkov resonances. }
\label{skr3c}
\end{figure}

To model  Cherenkov  radiation  it is often sufficient to
include third order dispersion only, so that a
single zero of  GVD is taken into account
\begin{eqnarray}
\label{eq7}
&& i\partial_z A-\left({1\over 2!}\beta_2 \partial^2_t +{i\over 3!}\beta_3
 \partial^3_t\right)A=\\
 && \nonumber -\gamma |A|^2A-T\gamma A \partial_t|A|^2.
\end{eqnarray}
Linear dispersive solution of  Eq. (\ref{eq7}) is
$A=e^{ikz-i\delta t}$ with $k(\delta)={1\over 2}\beta_2\delta^2+{1\over 6}\beta_3\delta^3$.
The zero GVD point  is located at
$\delta_0=-\beta_2/\beta_3$.  The soliton existence condition is $\beta_2<0$ and hence
if, e.g., $\beta_3>0$, then the GVD is normal ($\partial_{\delta}^2k>0$)
towards higher frequencies $\delta>\delta_0$, see Figs. \ref{skr3b}(c).
The pulse spectrum entering this range is not able to propagate in the soliton regime, which
leads to the radiation emission.

Small amplitude Cherenkov radiation $F_{Ch}$ with frequency detuned far from the soliton
 obeys \cite{WAI1990,Karpman1993,Akhmediev1995,Skryabin2003,Biancalana2004}
\begin{equation}
\left(i\partial_z-{1\over 2!}\beta_2 \partial^2_t -{i\over 3!}\beta_3
 \partial^3_t \right)F_{Ch}\propto \partial_t^3A_s.\label{skrn1}
 \end{equation}
Unlike  the above mentioned
Airy waves, the resonant radiation is  phase matching dependent
and therefore it is narrow band.
Using Fourier expansion $F_{Ch}=\int_{-\infty}^{\infty}\epsilon_{Ch}(\delta)e^{ik(\delta)z-i\delta t}d\delta$,
taking Eq. (\ref{sol}), and calculating $z$ derivatives of the phases involved, we then
equate the wave numbers of the left and right hand sides of Eq. (\ref{skrn1}) to find
the wavenumber matching condition
\begin{equation}
\label{eq8}
k(\delta)=k_s(\delta)
\end{equation}
The roots of Eq. (\ref{eq8}) are the resonance frequencies
$\delta_{Ch}$.  $\beta_3>0$ ($\beta_3<0$) gives the blue (red)
shifted dispersive waves, see Figs. \ref{skr3b}, \ref{skr3c},
respectively.
Accounting for the higher order dispersions is effortless and may
lead to several roots of Eq. (\ref{eq8}) \cite{Genty2004a,Falk2005,Frosz2005a}. Energy of the emitted wave
(or waves) is drawn from the entire soliton spectrum resulting in
the adiabatic decay of the soliton \cite{Skryabin2003,Biancalana2004}.

In a typical   supercontinuum setting with $\beta_3>0$ the Raman
effect increases frequency detuning ($\Delta=\delta_{Ch}-\delta_s$)
between the soliton  and its resonance radiation, see Fig.~\ref{skr3b}(c). For $\beta_3<0$ the
situation is the opposite, see Fig.~\ref{skr3c}(c). The amplitude of
the emitted radiation is proportional to $e^{-a^2 |\Delta|}$ ($a$ is
a constant), where $\Delta\propto z$
\cite{WAI1990,Karpman1993,Akhmediev1995,Skryabin2003,Biancalana2004}.
Thus for $\beta_3>0$ the soliton self-frequency shift induces
exponential decay of the radiation amplitude with propagation
\cite{Biancalana2004,Gorbach2006}. Practically this implies that the soliton
emits significant shortwavelength radiation only at the initial stage of the
supercontinuum generation. The radiation emission quickly becomes
unnoticeable when the solitons are shifted away from the zero GVD
point. Then the natural question is  what causes the continuous blue shift of the short-wavelength
edge of the supercontinuum, see Fig. \ref{skr2}.
The answer is - soliton-radiation interaction (not mere radiation emission)
\cite{Gorbach2007a,Gorbach2007c,Genty2004b,Genty2005,Cumberland2008,Travers2009,Stone2008,Kudlinski2008}.
The rest of this section and sections V and VI elaborate on this in great details.

\begin{figure}
\centering
\includegraphics[width=7cm]{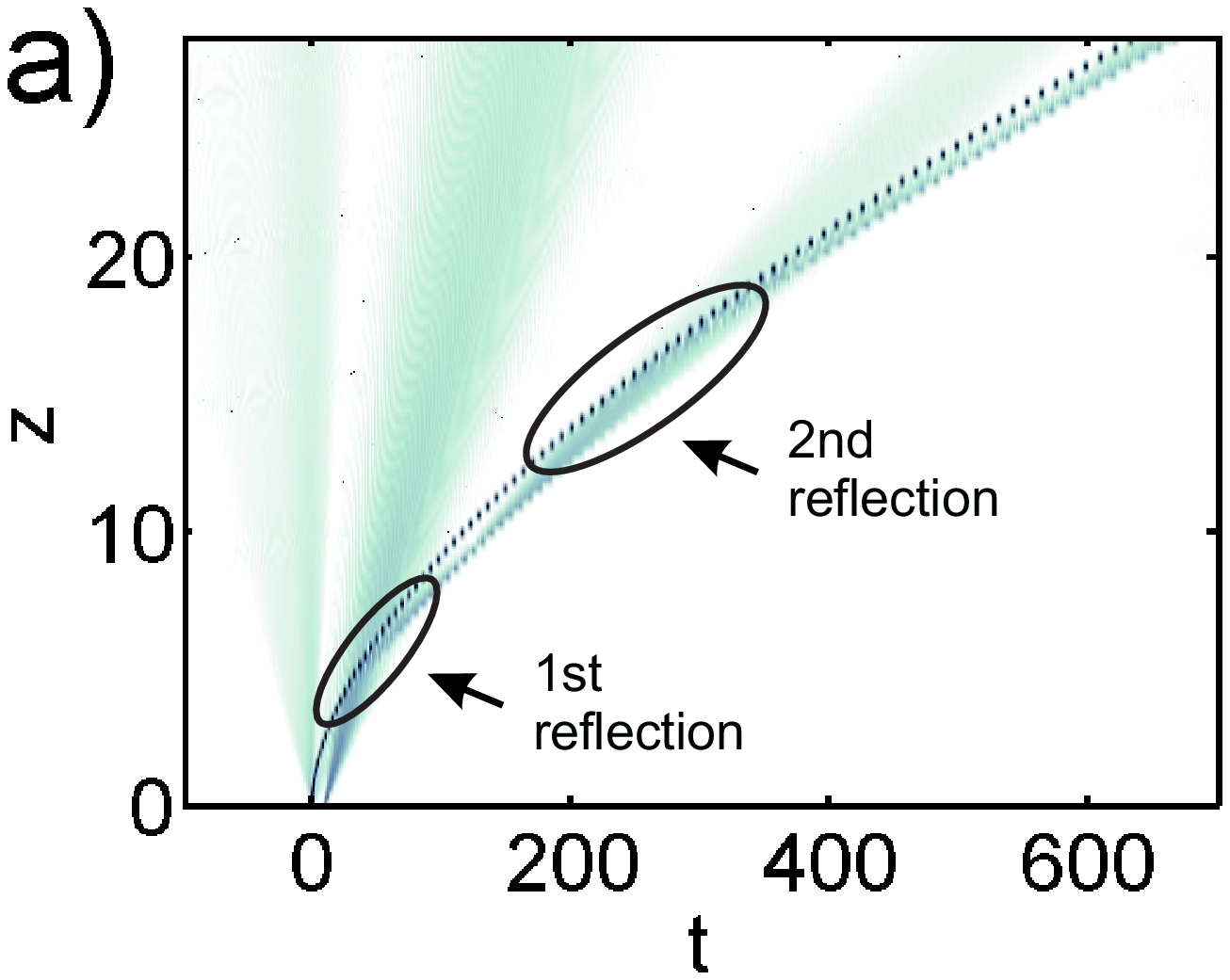}
\includegraphics[width=7cm]{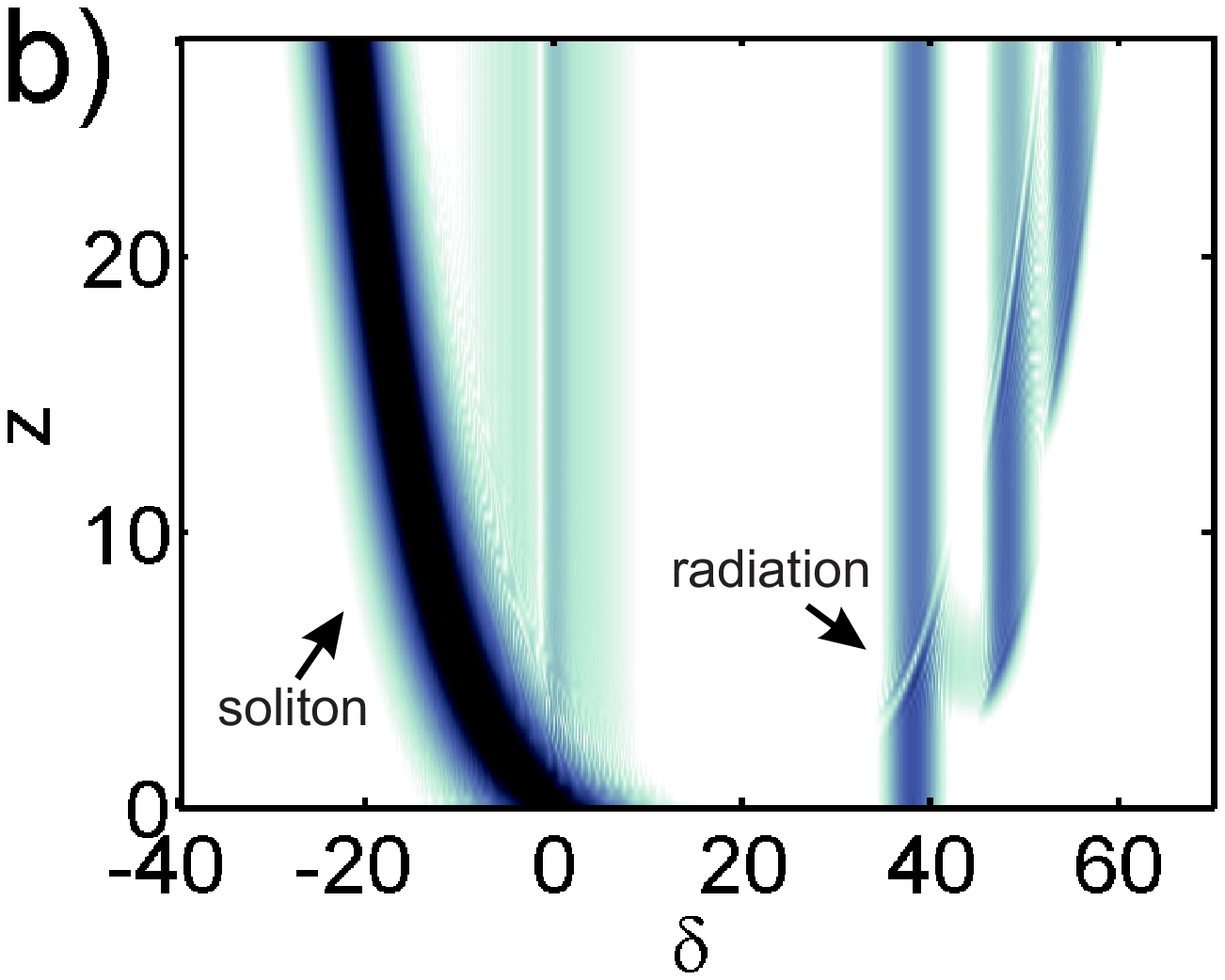}
\caption{A weak gaussian pulse in normal GVD range initially placed
behind the soliton experiences multiple reflections from the latter.
After each collision some light gets through the soliton and some is
reflected back. The reflected light is localized close to the
soliton and almost indistinguishable from it in the $(z,t)$-plot
(a). However, the reflected light is clearly visible in the spectral
$(z,\delta)$-plot (b), because its frequency is blue-shifted after
each collision.  } \label{skr3bis}
\end{figure}

\begin{figure}
\centering
\includegraphics[width=7cm]{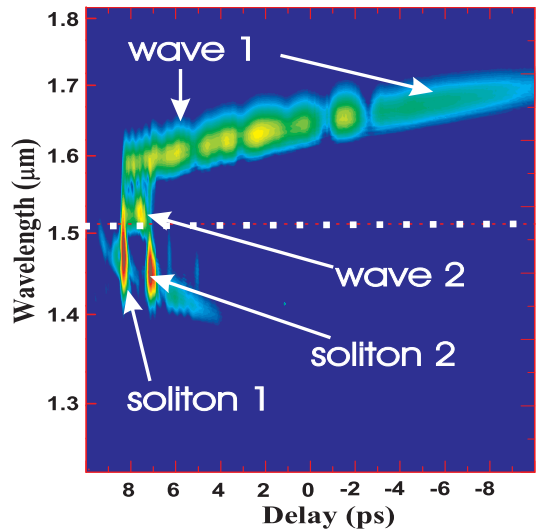}
\caption{(color online) Experimentally measured XFROG spectrogram showing emission of the strong long
wavelength radiation  and its reflection from the 2nd soliton in a fiber with $\beta_3<0$
\cite{Efimov2004}.
The strong resonant radiation (wave 1) is emitted by the soliton 1. The weaker soliton 2 reflects
a part of this wave backwards and simultaneously transforms frequency of the
reflected radiation (wave 2). Doted line marks the zero GVD wavelength.
}
\label{last}
\end{figure}

If $\beta_3>0$, then the resonant radiation emitted by the soliton has group velocity
less  than  the  soliton itself, so that the radiation
appears behind the soliton (Fig.~\ref{skr3b}(a)).
However, after some propagation the radiation catches up with the
soliton, which is continuously decelerated by the Raman effect,
and  the two  collide (Fig.~\ref{skr3b}(a), \ref{skr3bis}(a)).
During this collision the radiation is reflected from the soliton backwards,
and so the next collision becomes unavoidable, through the same mechanism.
The process can be repeated many times, see Fig. \ref{skr3bis}.
Note, that the radiation colliding with the soliton,
should not be necessarily emitted by the soliton itself,
it can  be initiated by other mechanisms, e.g., via SPM
or emitted by other solitons (Figs. \ref{skr3bis},\ref{last}).
An important condition for the reflection of the radiation from a soliton
to actually happen is that the radiation frequency should belong to the normal GVD range.
The radiation in the anomalous GVD range practically does not see solitons, since the model
is close to the ideal integrable NLS.

Reflection of the radiation backwards (towards smaller $z$ and larger $t$)
from the soliton, implies
that the radiation group velocity is further reduced, i.e. the group index
$\partial_{\delta}k$ for the radiation  increases.
For the normal GVD ($\partial_{\delta}^2k>0$) the increase in the group
index happens together with the increase in frequency $\delta$. Therefore the backward reflections
have to be accompanied by the blue shifting of the radiation frequency, see Fig. \ref{skr3bis}(b).
This is exactly  the type of process happening
between the solitons and the short wavelength radiation in the expanding supercontinuum shown
in Fig. \ref{skr2} \cite{Gorbach2006,Gorbach2007a,Gorbach2007c}. Obviously the reflection process
is nonlinear in its nature, which is explained in details in the next section.

For $\beta_3<0$, the detuning between the radiation and soliton gets smaller with propagation
(Fig. \ref{skr3c}(c)), therefore the radiation is exponentially amplified in $z$.
This amplification leads
to the strong spectral recoil on the soliton followed by the compensation of the soliton
self-frequency shift \cite{Skryabin2003,Biancalana2004,Efimov2004,Tsoy2006} (see Fig. \ref{skr3c}(b)).
In this case the radiation is emitted ahead of the soliton and
has no chance of interacting with it, see, Fig. \ref{skr3c}(a).
The radiation can, however, interact
with other solitons present in the fibers, bounce back from them and has its
frequency transformed \cite{Efimov2004,Gorbach2007c}, see Fig. \ref{last}.

We also note at the end of this section that dark  solitons existing for
the normal GVD are  known to emit Cherenkov radiation into
the anomalous GVD range \cite{Karpman1993a,Afanasjev1996}. Amplification of this radiation by the Raman effect
and its possible use for  generation of
broad continua has been recently investigated \cite{Milian2009}.

\section{Scattering of radiation  from  solitons and supercontinuum}

In its essence the scattering of a dispersive wave from a soliton
is a four-wave mixing (FWM) nonlinear process  sensitive to the phase matching conditions
\cite{Yulin2004,Skryabin2005,Efimov2004,Efimov2005a,Efimov2006,Gorbach2006}.
The latter, however, work out in an unusual way.
An important difference of the soliton-radiation interaction
with the text-book four-wave mixing of dispersive waves only \cite{Agrawal2001},
is that  one of the participating fields is a non-dispersing pulse (soliton) having straight-line
dispersion and moving with a group (not phase) velocity, see Eq. (\ref{eq6}).
\begin{figure}
\centering
\includegraphics[width=7cm]{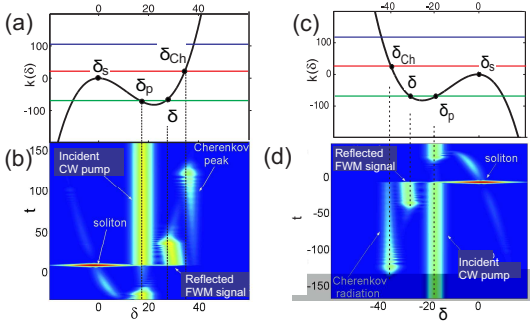}
\caption{(color online) Wave number matching diagrams (a,c) and XFROG spectrograms (b,d) for the fiber
pumped with the soliton and cw (Raman effect is disregarded).
(a,b) correspond to $\beta_3>0$ and (c,d) to $\beta_3<0$.
The black curved lines in (a,c) are $k=\beta_2\delta^2/2+\beta_3\delta^3/6$.
The upper  and lower  horizontal lines in (a,c) correspond to the FWM resonances
given by Eq. (\ref{eq13}) and (\ref{eq14}), respectively.
The middle horizontal  lines in (a,c) give Cherenkov resonances, Eq. (\ref{eq8}).}
\label{skr5}
\end{figure}

Lets assume that $F_p$ and $\delta_p$ are
the amplitude and frequency of the dispersive wave incident on the soliton $A_s$, while
$F$ is the reflected  signal field with an unknown frequency $\delta$.
The equation for  $F$ is  \cite{Skryabin2005}:
\begin{eqnarray}
\nonumber  \left(i\partial_z-{1\over 2!}\beta_2 \partial^2_t -{i\over 3!}\beta_3
 \partial^3_t \right)F= &&-\gamma  |A_s|^2F_p\\
  &&-\gamma A_s^2F_p^*~.\label{eq9}
\end{eqnarray}
Thus $F$ is exited by  $|A_s|^2F_p$ and $A_s^2F_p^*$ with relative efficiency of the
two excitation channels determined by the phase matching.
Assuming that the incident wave is
$F_p=\epsilon_pe^{ik(\delta_p)-i\delta_pt}$ and using Eqs. (\ref{eq3})-(\ref{eq6}),
we find the spectral content of the FWM terms:
\begin{eqnarray}
\nonumber && |A_s|^2F_{p}=\epsilon_{p}e^{ik(\delta_p)z}\times \\
\label{eq11} &&
\int_{-\infty}^{\infty}  f(\delta-\delta_p)e^{it_s[\delta-\delta_p]-i\delta t}d\delta~, \\
\nonumber  && A_s^2F_{p}^*=\epsilon_{p}^*e^{2i\phi(z)-ik(\delta_p)z}\times \\ && \int_{-\infty}^{\infty}
 f(\delta+\delta_p-2\delta_s)
e^{it_s[\delta+\delta_p-2\delta_s]-i\delta t}d\delta~,\label{eq12}
\end{eqnarray}
where  $f(\delta)=\int_{-\infty}^{\infty} \psi^2(t)e^{i\delta t}dt$.

Using Fourier expansion $F=\int_{-\infty}^{\infty}\epsilon(\delta)e^{ik(\delta)z-i\delta t}d\delta$
and taking $z$ derivatives of the phases involved (cf. Eq. (\ref{eq6}))
we equal the  wavenumber of $F$ to the wavenumbers of (\ref{eq11}) and (\ref{eq12}).
The result is
\cite{Yulin2004,Skryabin2005,Efimov2004,Efimov2005a,Efimov2006,Gorbach2006}
\begin{eqnarray}
\label{eq13} && k(\delta)=k_s(\delta)-[k_s(\delta_p)-k(\delta_p)],\\
\label{eq14} && k(\delta)=k_s(\delta)+[k_s(\delta_p)-k(\delta_p)].
\end{eqnarray}
Here $k$ is the wavenumber of the generated wave.
$k_s(\delta)$ and $k_s(\delta_p)$ are the soliton wavenumbers at the generated
frequency and at the frequency of the wave incident on the soliton, respectively.
For $\delta_s=0$,  $k_s(\delta)=q$
and  Eqs. (\ref{eq13}, \ref{eq14}) are simplified to the form
$k(\delta)=k(\delta_p)$, $k(\delta)=2q-k(\delta_p)$.

Solving Eqs. (\ref{eq13}), (\ref{eq14}) graphically we find that
they predict upto 4 resonances, see Fig. \ref{skr5}(a,c)  \cite{Skryabin2005}.
One  resonance is obvious $\delta=\delta_p$ and it  coincides with the frequency of the incident wave
(cw pump). $|A_s|^2F_p$ term and Eq. (\ref{eq13}) are responsible for two nontrivial resonances
falling into the regions of normal and anomalous GVD. The former one is typically much stronger
and corresponds
to the reflection of the wave from the soliton potential. The remaining one and the
resonance predicted by Eq. (\ref{eq14}) usually do not scatter much of the incident wave
and produce weak, but detectable signals.
Figs. \ref{skr5}(a,b) show the case when the incident
wave is reflected backwards from the soliton with simultaneous upshift of the reflected wave frequency.
Figs. \ref{skr5}(c,d) show the case when the radiation-soliton collision happens at the front
edge of the soliton and the wave is reflected ahead of the soliton.
The frequency is down shifted in the latter example.

\begin{figure}
\centering
\includegraphics[width=7cm]{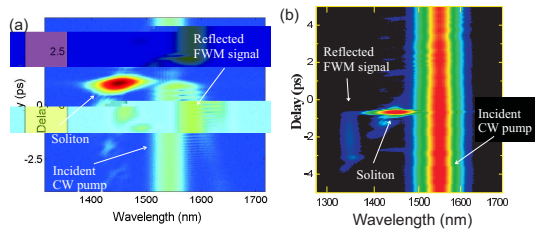}
\caption{(color online) Experimentally measured XFROG spectrograms showing radiation-soliton interaction.
(a)  corresponds to the  resonance
given by Eq. (\ref{eq13}) \cite{Efimov2005a} and (b) to  Eq. (\ref{eq14}) \cite{Efimov2006}.
(a)  shows the case of the forward reflected wave, as in Figs. (\ref{skr5})(d).
Both (a) and (b) measurements have been taken for $\beta_3<0$.}
\label{skr6}
\end{figure}

Eqs. (\ref{eq13}, \ref{eq14}) apply without
a change if the soliton and dispersive waves are orthogonally
polarized, which has been used in the  experimental measurements of the  soliton-radiation
interaction \cite{Efimov2005a,Efimov2006}. These measurements
fully confirmed validity of both
Eq. (\ref{eq13}) \cite{Efimov2005a} and Eq. (\ref{eq14}) \cite{Efimov2006}.
Corresponding examples of the experimental XFROG
spectrograms are shown in Fig. (\ref{skr6}).

It has been  verified that the  frequency up-shift of the
radiation,   resulting from the cascaded back reflection of the
radiation from intensity of the accelerating soliton $|A_s|^2$,
is the mechanism ensuring blue
shift of the short wavelength edge of the supercontinuum \cite{Gorbach2006}.
Phase matching conditions (\ref{eq13}) work out in a way  that with every reflection
the frequencies of the incident and reflected waves both
tend towards the limit point, where the group velocity of the dispersive wave
coincides with the soliton group velocity \cite{Skryabin2005,Gorbach2006}, see
Fig. \ref{skr6_1}. Indeed, if the instantaneous
soliton frequency is assumed to be zero, then  the minimum of $k(\delta)$
corresponds to the frequency with the group velocity matched to the soliton group velocity,
Fig. \ref{skr5}(a).
If the frequency of the incident wave ($\delta_p$)
is exactly at this minimum or very close to it, then
the resonance predicted by Eq. (\ref{eq13}) is either
degenerate or close to the degeneracy with $\delta_p$.
Under such conditions the process of frequency conversion practically happens
within a single wave packet and can be  termed as the
intrapulse four-wave mixing \cite{Gorbach2006}.
\begin{figure}
\centering
\caption{(color online) Spectral evolution along the fiber length. (a)-experiment and (c)-modelling.
(b) XFROG diagram  computed for the propagation distance $z=2$m.
(d) Group index as function of wavelength. The decrease in height of the shaded triangles
shows the decreasing frequency difference between the incident and reflected waves
\cite{Gorbach2006}.}
\label{skr6_1}
\end{figure}

In the above picture of the soliton-radiation interaction,
it remains unclear what sustains the efficiency of the process
over the long propagation distances.
Strong normal  GVD acting on the dispersive wave packet, should lead to
its sufficiently fast spreading and drop in the peak intensity terminating any nonlinear
interaction. Instead, as it has been originally measured in \cite{Hori2004}
and can be seen from the modelling results (Figs. \ref{skr2}(b), \ref{skr6_1}(b)),
the wavepackets on the  short-wavelength edge of the continuum remain
localised on the femtosecond time scale and propagate
in the soliton-like regime albeit in the normal GVD range.
This problem is addressed in the next section.

\section{Radiation trapping and supercontinuum}
So, what does stop dispersive spreading of the radiation wavepackets at the short wavelength edge
of supercontinuum
and keeps them localised on the femtosecond time scale? The short answer is - the refractive index change
created by the decelerated soliton  exerts a special type of inertial force
and ensures dispersionless propagation (trapping) of the radiation.

Reflection of  radiation from a soliton plays the major role in the trapping mechanism.
Mathematically this is described by  Eq. (\ref{eq9}) with the first nonlinear term in the
right-hand side, see the previous section.
Switching into the reference frame moving together with the soliton
reveals two distinct propagation regimes. If the offset of group velocities
of the soliton and radiation is sufficiently large and fiber length is relatively short,
then the reflection from the soliton
of course happens, but recurrent collisions leading to
trapping can be disregarded \cite{Skryabin2005}.
The trapping phenomenon becomes the dominant feature of the propagation, when group velocities
of the soliton and radiation are sufficiently close
\cite{Gorbach2007a,Gorbach2007c,Gorbach2007b,Stone2008,Travers2009a,Hill:09}.

Formal consideration of the problem
starts from  Eq. (\ref{eq9}) complemented by the equation for the soliton field.
We choose the reference frequencies of the soliton ($\delta_s$)
and radiation ($\delta_p$)  so that  the group velocities are matched ($k'(\delta_s)=k'(\delta_p)$)
across the zero GVD point and assume
\begin{equation}
\label{eq15}
E=A_s\exp[ik(\delta_s)z-i\delta_s t]+F \exp [ik(\delta_p)z-i\delta_p t].
\end{equation}
For the amplitudes $A_s(z,t)$ and $F(z,t)$ we make substitutions \cite{Gorbach2007a,Gorbach2007c}
\begin{eqnarray}
\label{eq16}
A_s=\psi(z,\xi)\exp\left[-it{ gz\over k''_s}+iqz+{1\over 3k''_s}g^2z^3\right]~,&&\\
\label{eq17}  F=\phi(z,\xi)\exp\left[-it{ gz\over k''_p}+i\lambda z+{1\over 3k''_p} g^2z^3\right]~,&&\\
~t_s= gz^2/2~, \xi=t-t_s~,
\nonumber k''_{s,p}=\partial_{\delta}^2k(\delta_{s,p})~.&&
\end{eqnarray}
We have introduced $g$  parameter to indicate that the deceleration rate for the
soliton interacting with radiation can be different from
$g_0$ for a pure soliton, see Eq. (\ref{eq4a}).
GVD for the soliton and radiation are anomalous ($k''_{s}<0$) and normal ($k''_{p}>0$), respectively.
Thus directions of the frequency shifts acquired by the two waves are  opposite,
see the first terms in the exponential factors of Eqs. (\ref{eq16}), (\ref{eq17}). The  group
indices felt by the soliton and radiation ($\partial_{z}t_s=gz$) increase with the same rate
(see also discussion in the previous section) and  hence  they experience equal negative acceleration.

\begin{figure}
\centering
\includegraphics[width=4cm]{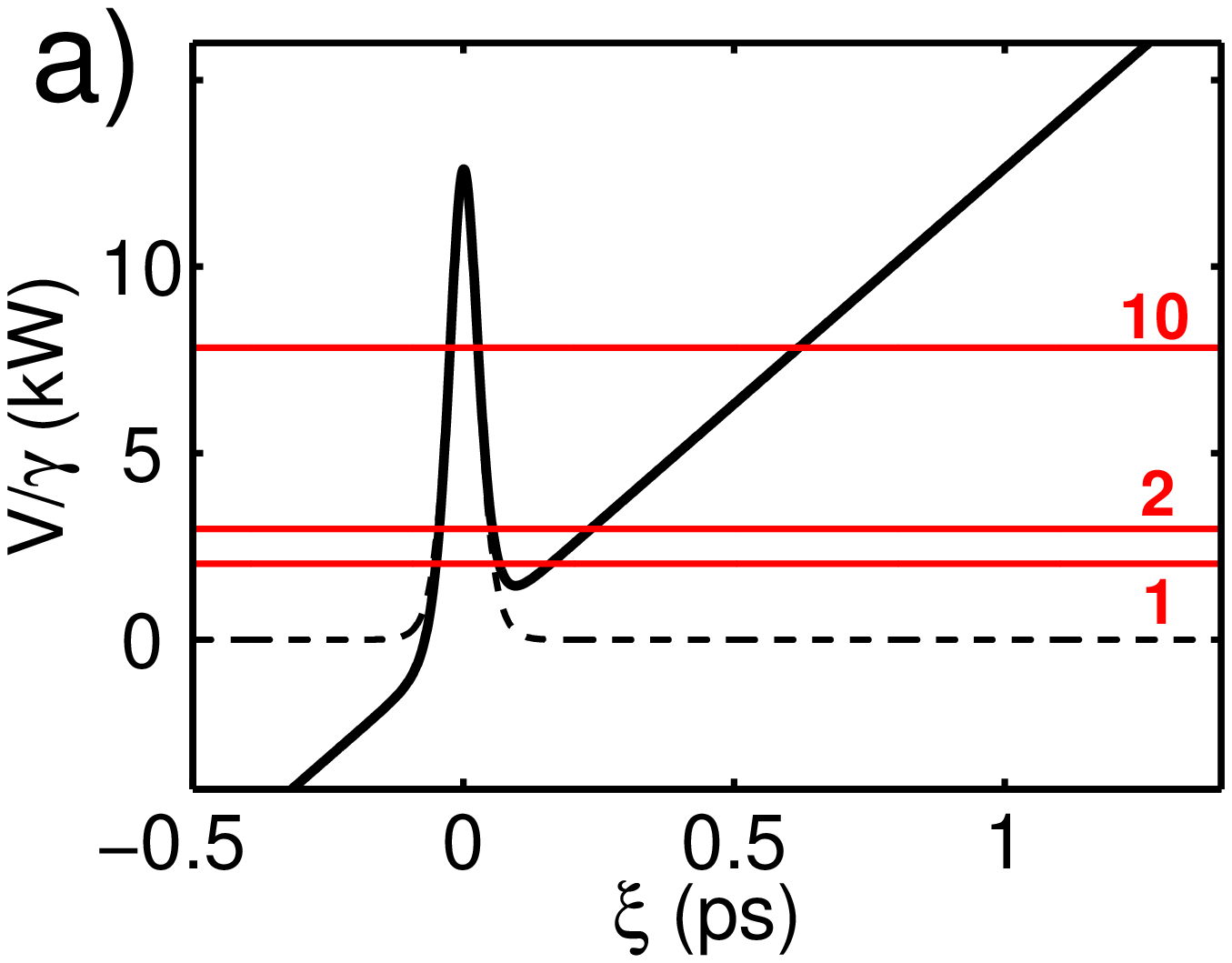}
\includegraphics[width=4cm]{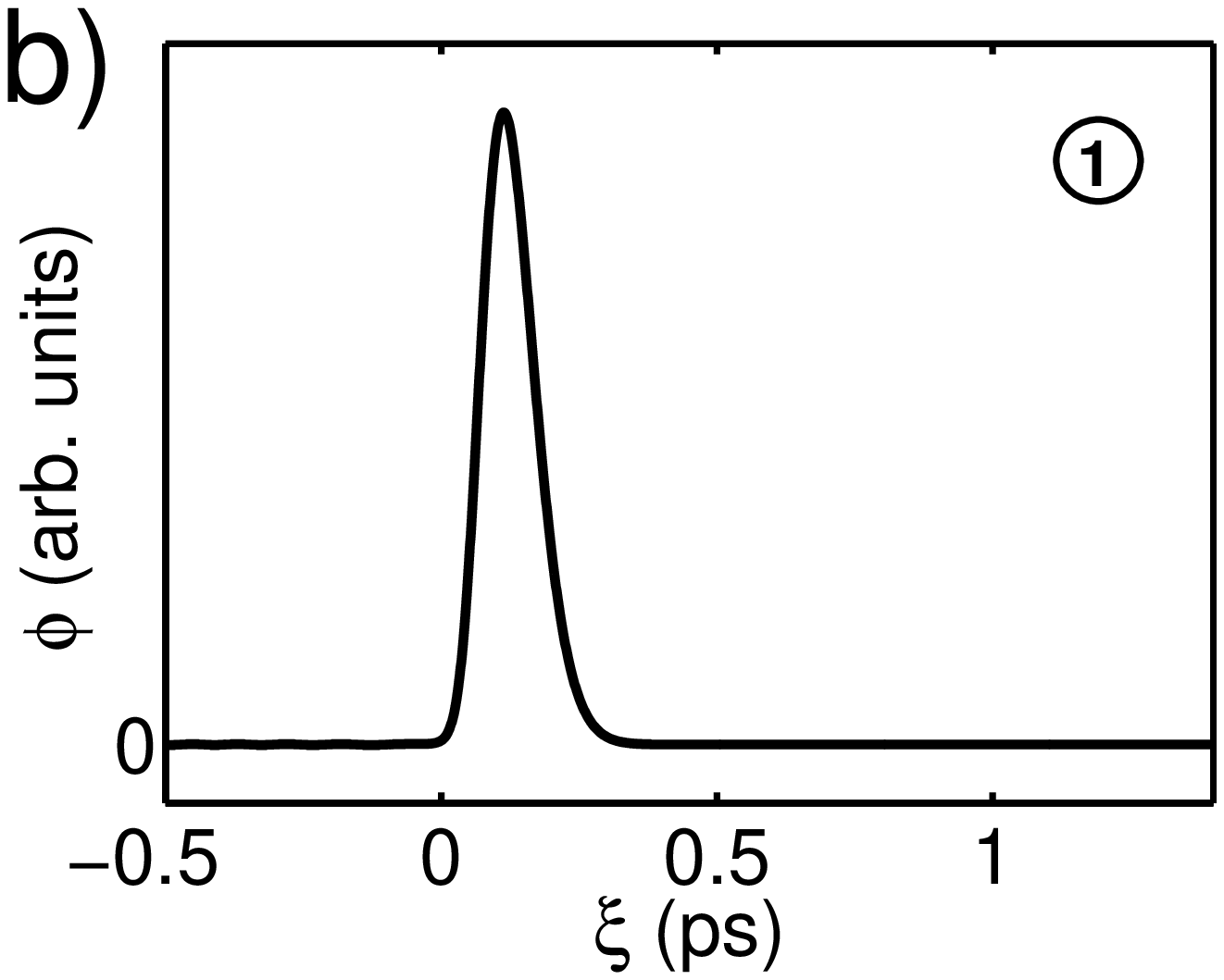}
\includegraphics[width=4cm]{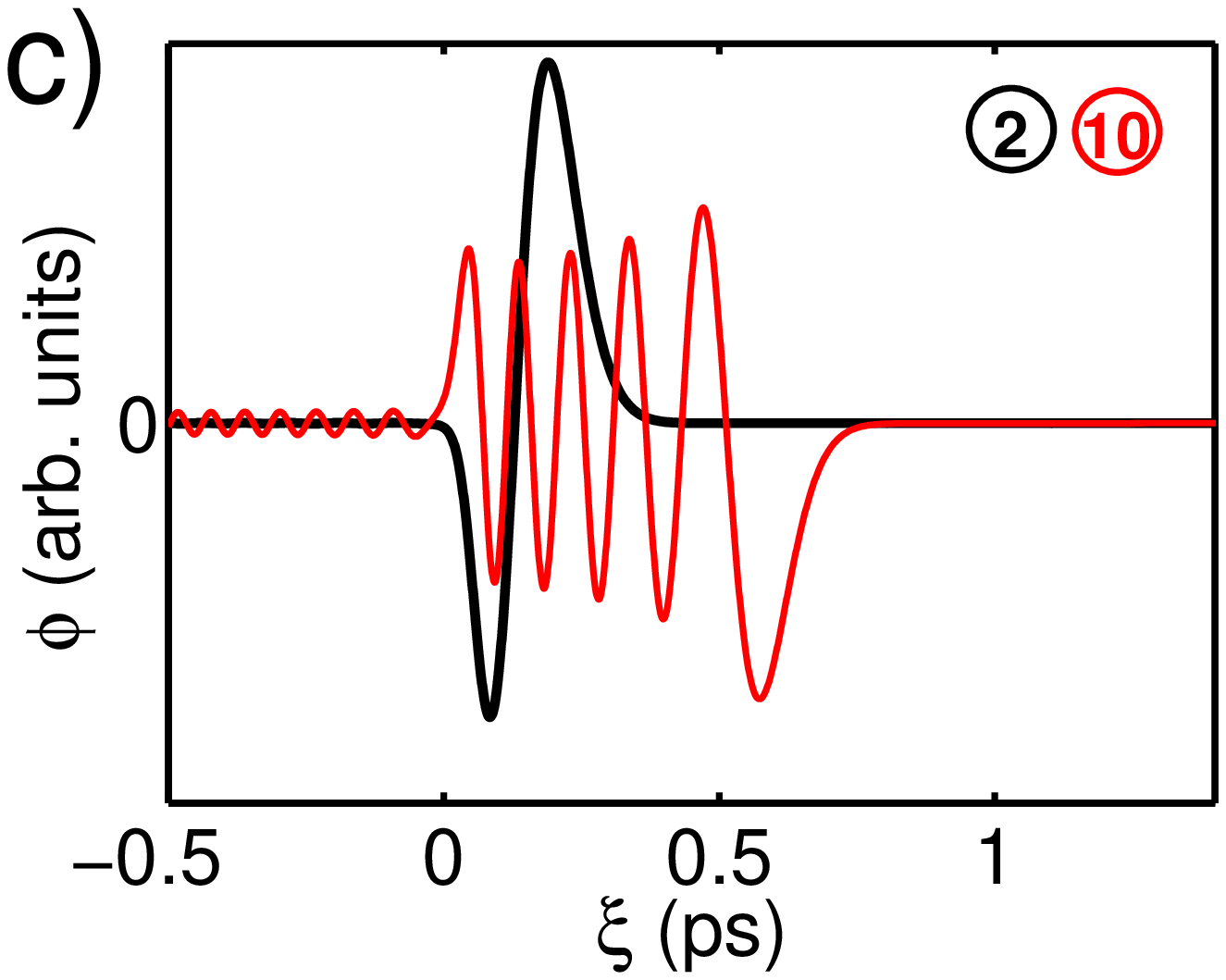}
\caption{(color online) (a) Full line shows the potential $V$. The dashed line shows the potential with the contribution from
the inertial force disregarded, which removes the  minimum and makes the radiation trapping impossible. Level lines
indicate ground state, 2nd and 10th modes.
(b) shows the ground state mode of the potential $V$ c) shows 2nd and 10th modes.}
\label{skr7}
\end{figure}

Now our problem is reduced to a formal question:
is there a solution for $\phi$ retaining its localized form?
In order to answer this formally we substitute (\ref{eq15}-\ref{eq17}) into Eq. (1)
and neglect several small terms, including the ones nonlinear in the radiation amplitude
$|\phi|^2$ \cite{Gorbach2007c}.
Then, the $z$-independent, i.e. shape preserving, radiation waves have to satisfy
\begin{equation}
\label{eq20}
 -{1\over 2}k''_p \partial^2_\xi \phi_n  +2\gamma|\psi|^2\phi_n
 +\frac{ g\xi}{k''_p}\phi_n=\lambda_n \phi_n~,k''_{p}>0~.
\end{equation}
The above is the linear Schr\"odinger equation with the effective potential energy
$V=2\gamma|\psi|^2+\frac{ g\xi}{k''_p}$. The first term inside $V$ is the  repelling potential
created by the refractive index change induced by the soliton. The dispersive wave reflects from it,
as described in the previous section. The second term in $V$ is the potential linearly increasing in $\xi$,
which exists only due to the fact that we have switched into the non-inertial frame of reference
accelerating together with the soliton. Hence this term represents a type of inertial force acting on photons.
It is  known from  classical mechanics,
that inertial forces act as usual ones, but show up in the equations of motion only,
when an appropriate non-inertial frame of reference is introduced.

\begin{figure}
\centering
\caption{(color online) Spectral evolution of the soliton-radiation bound states
with $n=1$ (a) and $n=3$ (b) given by Eqs. (\ref{eq16}),
(\ref{eq17}).} \label{skr8}
\end{figure}

Overall the potential $V(\xi)$ has the well defined minimum and therefore supports
quasi-localised modes (bound states), see Fig. \ref{skr7}. These modes can be found either numerically or
using a variational approach \cite{Gorbach2007c}.
Taking  soliton plus one of these modes and substituting them into Eqs. (\ref{eq16},\ref{eq17}) results in the
spectral evolution shown in Fig. \ref{skr8}.
The soliton spectrum moves continuously to the smaller frequencies and
the radiation spectrum moves towards higher frequencies.
The sign of $\beta_3$ is implicit, but very important here.
If $\beta_3>0$, as in the typical supercntinuum generation experiments,
then initially $\delta_s<\delta_p$ and radiation and soliton spectrally diverge with propagation.
However, if $\beta_3<0$, then $\delta_s>\delta_p$, while the spectral shifts still act in the
same directions, so that frequencies of the radiation and soliton converge with propagation \cite{Gorbach2007c}.

\begin{figure}
\centering
\includegraphics[width=7cm]{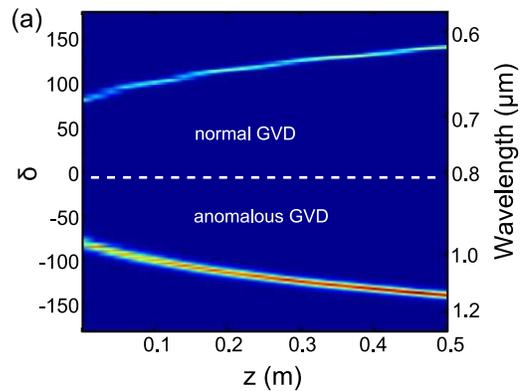}
\includegraphics[width=7cm]{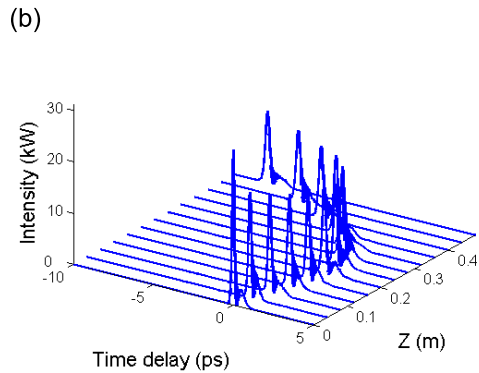}
\caption{(color online) Spectral (a) and time-domain (b) evolution of the soliton
and trapped radiation modeled using the realistic frequency
dependence of GVD (cf. (a) with Fig. \ref{skr8}). (b) is plotted in
the reference frame accelerating with $g_0$, which explains the
curvature of the soliton trajectory opposite to the one in Fig.
\ref{skr3b}(a).} \label{skr8a}
\end{figure}

One can notice that taking the higher
order modes leads  to temporal (Fig. \ref{skr7}(c)) and
spectral (Fig. \ref{skr8}) broadening of the radiation.
Spectral trajectories in Fig. \ref{skr8} follow the straight lines
because Eqs. (\ref{eq16}), (\ref{eq17}) assume frequency independent GVD.
In a real fiber the soliton and radiation moving away from the zero GVD point
encounter increasing absolute values of the GVD. This leads to the adiabatic broadening of the
soliton and  reduces  its Raman shift (Fig. \ref{skr8a}(b)).
The result is the gradually slowing spectral divergence of the soliton and radiation (Fig. \ref{skr8a}(a)).
Physically, the frequency conversion of the radiation wavepacket
is due to the intrapulse four-wave mixing (see previous section),
which is made possible by the sustained overlap of the radiation and soliton pulses.

Each of the solitons inside the supercontinuum shown in Figs.
\ref{skr1}(b), \ref{skr6_1} has its own radiation pulse continuously
drifting towards shorter wavelengths. We have found that the
strongest soliton on the long-wavelength edge of the supercontinuum
spectrum in Fig. 1(a) creates the potential $V$ trapping around 20
modes on the short-wavelength edge. The radiation captured by the
soliton  can be represented as a superposition of these modes.
Adiabatic transformation of the soliton power and width with
propagation, caused by the increasing dispersion, induces weak
adiabatic evolution of the mode parameters, but apart from this the
modes are stationary solutions and hence their temporal and spectral
dynamics are suppressed. Therefore approximating the radiation field as
\begin{equation}
F(z,\xi)=e^{-it gz/k''_2+ g^2z^3/(3k''_2)}
\sum_n\phi_n(\xi)e^{i\lambda_n z}\label{mode}\end{equation}
gives the very good matching with numerically computed spectral
evolution of the short wavelength edge of the supercontinuum
\cite{Gorbach2007a}.

At the end of this section we'd like to draw your attention to few
points. Firstly, the potential barrier, see Fig. (\ref{skr7}), on the soliton side is high,
but still finite, so that some light leaks through it. The leaked radiation is
especially noticeable in the higher-order modes, Figs.
\ref{skr1}(b), \ref{skr7}(c). Secondly, from numerical modelling of
supercontinuum it can be found that the rate of the soliton
self-frequency shift on the red edge of the supercontinuum is
actually higher than for the bare soliton. Analytically this effect
can be captured if $g$ is calculated with  nonlinear in $\phi$ terms
accounted for \cite{Gorbach2007c}. The resulting expression is $g= g_0
(1+Pb^2)$. Here $P$ is the peak power of the radiation and $b^2$ is
a constant.
Thirdly, to generate spectra shown in
Figs. \ref{skr2} and \ref{skr6_1}(a), (c) we have used the same
dispersion profiles and input powers, but different input
wavelengths. For the Fig.  \ref{skr6_1}(a), (c) the pump was only
$10$nm away from the zero GVD point, this has led to formation of
less powerful solitons. Hence their frequency shift and associated
frequency shift of the shortwavelength radiation were substantially
smaller leading to much narrower continua. Fourthly,  it has been
reported that for high pump powers the solitons at the infrared edge
of the continuum tend to collide and  form bound states
\cite{Podlipensky2008}. These effects are naturally expected to
influence radiation at the short wavelength edge, through the change
of the trapping potential and the collision induced changes in soliton frequencies
\cite{Luan2006a}.

It is important to note, that the simultaneous and opposite
soliton-radiation frequency conversion
and radiation trapping have been observed in few experiments not related to the
mainstream of the fiber supercontinuum research.
First known to us paper is the 1987 experiment by Beaud et al  \cite{Beaud1987}.
Then there was a gap until 2001, when Nishizawa and Goto reported
a series of  spectral and time domain measurements of the effect of pulse
trapping by a soliton across the zero GVD point \cite{Nishizawa2001,Nishizawa2002}.
For recent experimental observations and frequency conversion
applications of the trapping effect, see, e.g.
\cite{Cumberland2008,Stone2008,Kudlinski2008,Hill:09,Nishizawa2009}

\section{Gravity-like effects,  freak waves and turbulence of light in fiber}
Parameter $g$ used in the above section has an obvious analogy with the  acceleration of free fall.
Eq. (\ref{eq20}) can be interpreted as the equation for a quantum particle
in the  gravity field  and subject to the additional potential created by the soliton.
If the radiation wave is prepared in such a way that it
is both well localised in time and shifted away from the potential minimum,
 then it is a highly multimode state.
Thus, according to the correspondence principle, one should expect quasi-classical dynamics to be seen.
The  wave packet should roll down a linear
potential towards the soliton, reflect back from the latter  and, after some time, reconstruct
itself in the original location. It is known though, that the quasi-classical bouncing carries on only for
limited  time, until it is replaced by the complete delocalization of the wave packet,
which restores  its shape again later. This effect is known as quantum bouncing \cite{Robinett2004}.
Previously bouncing has  been observed with clouds of Bose-condensed
ultracold atoms subject to the field of gravity and reflecting of an atom mirror, see
\cite{Bongs1999,Saba1999} and Fig. \ref{skr100}(a).
Fig. \ref{skr100}(b) shows numerically modeled space-time evolution of the radiation pulse
in the normal GVD range of an optical fiber bouncing on a decelerating  soliton \cite{Gorbach2007b}.
Similar light bouncing effects have also been reported for the curved
waveguide arrays \cite{Longhi2008,DellaValle2009}.
An important feature of our case, is that on each reflection from the soliton the
frequency of the radiation is up shifted (see Fig. \ref{skr100}(c)).
Recently, reflection of the radiation from the soliton and the associated blue shift have been interpreted
as the frequency shift  at the white-hole horizon \cite{Philbin2008}.
The same work has
predicted that the quantum effects of horizons, in particular Hawking radiation,
can potentially be seen due to soliton-radiation interaction in optical fibers.
\begin{figure}
\centering
\caption{(color online) (a) Series of images of the condensate bouncing off a light sheet \cite{Bongs1999}.
Time interval between the images is $2$ms along the horizontal axis.
(b) Numerical modeling of the dispersive radiation bouncing off the decelerating soliton.
$\xi$ is the dimensinless time measured in the reference frame moving together with the soliton
and $z$ is the distance along the fiber
(c) frequency, $\delta$, conversion of the bouncing radiation. \cite{Gorbach2007b}} \label{skr100}
\end{figure}

Another problem recently posed by
the supercontinuum and soliton research has been the question
about existence of optical freak or rogue waves \cite{Solli2007}.
This phenomenon has been actively studied in the context of ocean waves,
where the rare waves, with probability not described by the tails of the gaussian distribution, and the
amplitude few times larger than the average (for the current conditions) wave hight
pose serious and hardly predictable threat for  ships and offshore industries.
NLS model is known to describe deep water waves including the freak events \cite{Janssen2003}.
Therefore one can expect appearance of similar phenomena in fiber optics.
In particular,  cases of the notable pulse to pulse fluctuations of the supercontinumm,
 can  be attributed to generation of the infrared solitons with
unusually large amplitudes \cite{Solli2007,Dudley2008,Solli2008}.
Probability of this to happen is described by the tail of the L-shaped distribution function.
These  freak solitons emerge essentially due to anomalously
strong focusing developing in the course of modulational instability and the higher order soliton fission
\cite{Solli2007,Dudley2008,Solli2008}.
There is also another class of localised freak wave solutions of NLS equation. These solution are
breathers, i.e. localized waves periodically absorbing and releasing their energy into
the continuous wave background \cite{Akhmediev2009,AKHMEDIEV1992}.
This type of waves has the property of sudden appearance and disappearance,
which is a known feature of the oceanic freak waves, but still
has not been seen in optical fibers.

Another area, where analogy between  fluid mechanics and fiber optics is starting to produce
interesting results is the turbulence research \cite{DYACHENKO1992,Barviau2008,Barviau2009}.
Spectral broadening due to multiple four-wave mixing processes of random
weakly nonlinear waves can be associated with irreversible evolution
of the spectrally narrow pump towards spectrally broad thermodynamic equilibrium
\cite{Barviau2008,Barviau:09,Barviau2009}.
Current theoretical approaches to the turbulence in general and to the turbulent
supercontinuum  still have not progressed to the level where the
role of mixing of incoherent wave fields with
coherent wave structures (solitons) in spectral broadening
can be fully revealed \cite{DYACHENKO1992}.
At the same time fiber supercontinuum research
suggests that this difficult case is the most practically relevant.
Complex 'far from equilibrium'  dynamics is also well known in
the fiber based systems, such as fiber lasers and coherently pumped
fiber resonators, where soliton pulses and spectral
broadening  often coexist, see, e.g.,
\cite{Mitschke1996,Lee2005,Babin2008,Peng2008,kozyreff:043905,Chouli2009}.

\section{Summary}
For convenience of our readers we summarize here those of the  soliton properties,
which are most important for supercontinumm generation
\begin{itemize}
\item Interaction of a soliton with dispersive radiation leads to the phase-matching
sensitive generation of new frequencies. The most pronounced, out of few possible
interaction channels, is the reflection of the radiation from the
refractive index change created by the soliton intensity.
The reflection happens providing the radiation frequency belongs to the normal GVD range.
Depending on the sign of the 3rd order dispersion and frequency of the incident radiation,
the frequency of the reflected wave gets  either up- or down-shifted.

\item  Raman effect  decelerates solitons and down-shifts their frequency.
Such solitons can interact with radiation repeatedly,
trap it on the time scales of 100fs and continuously up-shift
the radiation frequency.
\end{itemize}
The prevalent scenario of the supercontinuum generation in
photonic crystal fibers pumped by femtosecond pulses with the input wavelength around the
zero GVD point can be summarized as:
\begin{itemize}
\item Spectrum of the input pulse is distributed in some proportion between the frequency ranges
with normal and anomalous GVD. This happens through the combination of nonlinear processes.
SPM dominates during the first few centimeters of propagation.
Then the soliton fission  accompanied by the
radiation emission and reflection of the dispersive waves from emerging
solitons lead to further spectral broadening.
\item The next stage is when the Raman shifted solitons on the long-wavelength edge of the supercontinuum
enter into the regime of the cascaded  interaction with dispersive radiation. This quickly leads to
formation of the bound soliton-radiation states responsible for continuous spectral
divergence of the supercontinuum edges. The necessary condition for this to happen
is the near matching of the group velocities across the zero GVD point.
\end{itemize}
Using nano-second or cw pump  for supercontinuum generation leads to modulational instability and
subsequent creation of a soliton train. The latter traps radiation and the above
scenario is realized again albeit with greater number of solitons
\cite{Cumberland2008,Travers2009,Kudlinski2008,Stone2008}
and greater sensitivity to noise \cite{Solli2007,Turitsyn2008}.

\section{Perspectives}
Interaction between solitons and radiation, optical turbulence,
freak waves, and development of ideas
around the gravity-like forces exerted on light by solitons
all are on the list of problems stimulated by the fiber
supercontinuum research and undergoing the stage of active exploration.
Supercontinuum generation has been of course known outside the
fiber context in bulk solids, liquids and gases, see, e.g. \cite{Kolesik2008,Berge2007,Couairon2007},
where all three space dimensions are important. Role of spatial and spatio-temporal solitons
in these systems is still far from been fully explored \cite{Yulin2005,Berge2007}.
Generation of broad spatial and frequency spectra in nonlinear photonic crystals
is another area where interaction of solitons with diffracting and dispersing waves
can be important \cite{Bartal2006,Manela2006,Jia2007,Babushkin2007,Dong2008,Benton2008}.
Strong field localisation in metalic optical nano-antennas has been demonstrated to lead
to supercontinuum generation \cite{Muhlschlegel2005},
thereby  linking the effects discussed above with nanophotonics.
A possibility of nonlinear optical processes in nature made optical waveguides, e.g.,
found in  sea organisms and  sometimes having a structure of
fibers with few micron core diameters and photonic crystal cladding
\cite{Aizenberg2004,Kulchin2008},  remains an intriguing problem to consider.
Overall, a bidirectional flow of ideas between fiber photonics
and other branches of optics and physics in general
is more than likely to stimulate further progress in the soliton
and supercontinuum related research.

%\bibliographystyle{apsrmp}
%\bibliography{SOLandWJWandBG_sp}

\end{document}